%% file: paper-epsK.tex
\documentclass[prd,twocolumn,showpacs,showkeys,preprintnumbers,floatfix,
  nofootinbib]{revtex4-1}
\usepackage{amsfonts} 
\usepackage{amssymb} 
\usepackage{amsmath} 
\usepackage{subfigure}
\usepackage{graphicx} 
\usepackage{array} 
\usepackage{dcolumn} 
\usepackage{bm} 
\usepackage{latexsym} 
\usepackage{longtable} 
\usepackage{hyperref} 
\usepackage[usenames,dvipsnames]{xcolor}
\usepackage{mathrsfs}
\usepackage{comment}
\usepackage{rotating}
\usepackage{here}
\usepackage{float}
\usepackage{multirow}

\graphicspath{{./figs/}}

\allowdisplaybreaks

\input{macro.tex}
\begin{document}

\title{Standard Model evaluation of $\varepsilon_K$ using lattice QCD
  inputs for $\hat{B}_K$ and $V_{cb}$}

\author{Jon A. Bailey}
\affiliation{
  Lattice Gauge Theory Research Center, FPRD, and CTP, \\
  Department of Physics and Astronomy,
  Seoul National University, Seoul, 151-747, South Korea
}

\author{Yong-Chull Jang}
\affiliation{
  Lattice Gauge Theory Research Center, FPRD, and CTP, \\
  Department of Physics and Astronomy,
  Seoul National University, Seoul, 151-747, South Korea
}

\author{Weonjong Lee}
\email[E-mail: ]{wlee@snu.ac.kr}
%
%
\affiliation{
  Lattice Gauge Theory Research Center, FPRD, and CTP, \\
  Department of Physics and Astronomy,
  Seoul National University, Seoul, 151-747, South Korea
}

\author{Sungwoo Park}
\affiliation{
  Lattice Gauge Theory Research Center, FPRD, and CTP, \\
  Department of Physics and Astronomy,
  Seoul National University, Seoul, 151-747, South Korea
}

\collaboration{SWME Collaboration}

\date{\today}

\begin{abstract}
We report the Standard Model evaluation of the indirect CP violation
parameter $\varepsilon_K$ using inputs determined from lattice QCD:
the kaon bag parameter $\hat{B}_K$, $\xi_0$, $|V_{us}|$ from the
$K_{\ell 3}$ and $K_{\mu 2}$ decays, and $|V_{cb}|$ from the axial
current form factor for the exclusive decay $\bar{B} \to D^* \ell
\bar{\nu}$ at zero-recoil.
The theoretical expression for $\varepsilon_K$ is thoroughly reviewed to give
an estimate of the size of the neglected corrections, including
long distance effects.
The Wolfenstein parametrization $(|V_{cb}|, \lambda, \bar{\rho},
\bar{\eta})$ is adopted for CKM matrix elements which enter through
the short distance contribution of the box diagrams.
For the central value, we take the Unitarity Triangle apex
$(\bar{\rho}, \bar{\eta})$ from the angle-only fit of the UTfit
collaboration and use $V_{us}$ as an independent input to fix
$\lambda$.
We find that the Standard Model prediction of $\varepsilon_K$ with
exclusive $V_{cb}$ (lattice QCD results) is lower than the
experimental value by $3.4\sigma$.
However, with inclusive $V_{cb}$ (results of the heavy quark
expansion), there is no gap between the Standard Model prediction of
$\varepsilon_K$ and its experimental value.
For the calculation of $\varepsilon_K$, we perform the renormalization
group running to obtain $\eta_{cc}$ at next-to-next-to-leading-order;
we find $\eta_{cc}^\mathrm{NNLO}=1.72(27)$.
\end{abstract}

\maketitle

\section{Introduction}
\label{sec:intr}

CP violation in nature was first discovered in an experiment with the
neutral kaon system in 1964 \cite{Christenson1964}.
There are two kinds of CP violation in the neutral kaon system: one is
the indirect CP violation due to CP-asymmetric impurity in the kaon
eigenstates in nature, and the other is the direct CP violation due to
the CP violating nature of the weak interaction
\cite{AlaviHarati:1999xp, Fanti:1999nm}.
CP violating observables are prime candidates in searches for physics
beyond the Standard Model.
Experimentally, CP violation in the neutral kaon system is known more
precisely than in any other physical system.
Here, we focus on the indirect CP violation in neutral kaons.

Indirect CP violation in the neutral kaon system is parametrized by
$\epsK$
\begin{equation}
  \label{eq:epsK_def}
  \epsK 
  \equiv \frac{\mathcal{A}(K_L \to \pi\pi(I=0))} 
              {\mathcal{A}(K_S \to \pi\pi(I=0))} \,,
\end{equation}
where $K_L$ and $K_S$ are the neutral kaon states in nature, and
$I=0$ represents the isospin of the final two-pion state.
%
%
%
%
%
In experiment \cite{Agashe2014:ChinPhysC.38.090001},
\begin{align}
  \label{eq:epsK_exp}
  \epsK &= (2.228 \pm 0.011) \times 10^{-3} 
  \times e^{i\phi_\eps} \,,\CL
  \phi_\eps &= 43.52 \pm 0.05 {}^\circ \,.
\end{align}
Here, the $\epsK$ value represents an $\approx 0.2\%$ impurity of the
CP even eigenstate in the $K_L$ state, which contains 99.8\% of the
CP odd  eigenstate.

We can also calculate $\epsK$ directly from the Standard Model (SM).
In the SM, CP violation comes solely from a single phase in the CKM
matrix elements \cite{Kobayashi1973:ProgTheorPhys.49.652,
  Cabibbo1963:PhysRevLett.10.531}.
The SM allows the mixing of neutral kaons $K^0 (d\bar{s})$ and
$\wbar{K}^0 (s\bar{d})$ through loop processes, and describes
contributions to the mass splitting $\Delta M_K$ and $\epsK$.
Hence, we can test the SM through CP violation by comparing
the experimental and theoretical values of $\epsK$.

In the SM, the master formula for $\epsK$ is
\begin{align}
  \label{eq:epsK_SM_1}
  \epsK
  =& e^{i\theta} \sqrt{2}\sin{\theta} 
  \Big( C_{\eps} \hat{B}_{K} X_\text{SD}
  + \frac{ \xi_{0} }{ \sqrt{2} } + \xi_\text{LD} \Big) \CL
   &+ \mathcal{O}(\omega\eps^\prime)
   + \mathcal{O}(\xi_0 \Gamma_2/\Gamma_1) \,,
\end{align}
where $C_{\eps}$ is a well-known coupling given in
Eq.~\eqref{eq:C_eps}, and $X_\text{SD}$ is the short distance
contribution from the box diagrams given in Eq.~\eqref{eq:SDfactorX}.
Here, the major contribution to $\epsK$ comes from the $\BK$ term, and
the minor contribution of about 5\% comes from the $\xi_0$ term.
The remaining contribution of $\xi_\text{LD}$ is about 2\% coming from
the long distance effect on $\epsK$ \cite{
  Christ2012:PhysRevD.88.014508, Christ:2014qwa}.
In Section \ref{sec:rev}, we re-derive the leading contribution given
in Refs.~\cite{ Buras2008:PhysRevD.78.033005,
  Buras2010:PhysLettB.688.309, Donoghue1992:Dynamics}.
We also explicitly derive higher order corrections, including the
long distance contribution, in this paper.
A similar formula without the long distance correction $\xi_\text{LD}$
and higher order terms appears in Refs.~\cite{
  Buras2008:PhysRevD.78.033005, Buras2010:PhysLettB.688.309}.
%

In order to calculate $\epsK$ directly from the SM, we use input
parameters obtained from lattice QCD and experiments.
In particular, $\hat{B}_{K}$ and $V_{cb}$ have dominated the
statistical and systematic uncertainty in the SM evaluation of $\epsK$
for a long time.

During the past decade, lattice QCD has made significant progress in
calculating $\hat{B}_K$ so that its error is reduced dramatically,
down to the $\approx 1.3\%$ level at present.
This result is available from the Flavour Lattice Averaging Group
(FLAG) \cite{Aoki2013:hep-lat.1310.8555}.
It is obtained by taking an average of the $\hat{B}_K$ results from a
number of lattice QCD groups \cite{ Bae:2013lja, Arthur:2012opa,
  Laiho:2011np, Bae2012:PhysRevLett.109.041601,
  Durr2011:PhysLettB.705.477}.
We calculate $\epsK$ using two different input values of $\hat{B}_K$:
one is the FLAG result \cite{Aoki2013:hep-lat.1310.8555}, and the
other is the most updated result from the SWME collaboration
\cite{Bae2014:prd.89.074504}.

It is also noteworthy that the lattice calculation of the amplitude
$\mathrm{Im}A_2$ related to the decay $K \to \pi\pi (I=2)$
\cite{Blum2011:PhysRevLett.108.141601} makes it possible to determine
$\xi_0$ more precisely.

Another important input parameter to $\epsK$ is $V_{cb}$.
There are two independent methods to determine $V_{cb}$: one is the
exclusive method \cite{Bailey2014:PhysRevD.89.114504}, and the other
is the inclusive method \cite{Gambino2014:PhysRevD.89.014022,Alberti2014:PhysRevLett.114.061802}.
In the exclusive method \cite{Bailey2014:PhysRevD.89.114504}, one uses
lattice QCD to calculate semileptonic form factors for the decays
$\bar{B} \to D^{(*)} \ell \bar{\nu}$.
In the inclusive method \cite{Gambino2014:PhysRevD.89.014022,Alberti2014:PhysRevLett.114.061802}, one
performs analysis on $B \to X_c \ell \bar{\nu}$ decay processes using
the heavy quark expansion \cite{Uraltsev:2000qw}.
Here, we use both the exclusive and inclusive $V_{cb}$ to determine
$\eps_K$, and we compare the results with each other and experiment.
%

We use the Wolfenstein parametrization for the CKM matrix
\cite{Buchalla:1995vs}, truncating the series at
$\mathcal{O}(\lambda^7) \approx 10^{-5}$.
Here, we use three different choices of Wolfenstein parameters: (1)
$\lambda$, $\bar{\rho}$, and $\bar{\eta}$ from the global fit of the
CKMfitter collaboration \cite{ Charles:2004jd, Hocker:2001xe}, (2)
$\lambda$, $\bar{\rho}$, and $\bar{\eta}$ from the global fit of the
UTfit collaboration \cite{ Bona:2005vz, Bona:2007vi}, and (3)
$\bar{\rho}$ and $\bar{\eta}$ from an angle-only fit (AOF) from the
UTfit collaboration \cite{Bevan2013:npps241.89}, with an independent
input for $\lambda$ directly from $V_{us}$
\cite{Agashe2014:ChinPhysC.38.090001}.
In all the cases, we take $V_{cb}$ instead of the Wolfenstein
parameter $A$ from the unitarity triangle (UT) analysis.
We emphasize that the AOF does not use $\epsK$, $\hat{B}_K$, and
$V_{cb}$ to determine the UT apex $\bar{\rho}$ and $\bar{\eta}$.
Hence, it provides a self-consistent way to test the validity of the
SM with $\epsK$, using the lattice results for $\hat{B}_K$ and
$V_{cb}$ with no correlation between $(\hat{B}_K, V_{cb})$ and
$(\bar{\rho}, \bar{\eta})$.

To estimate the effect of correlations in lattice input parameters, we
note that $V_{cb}$ dominates the error in $\epsK$, and the FLAG
$\hat{B}_K$ \cite{Aoki2013:hep-lat.1310.8555} is dominated by the BMW
collaboration result \cite{Durr2011:PhysLettB.705.477}.
We assume that there is no correlation between the BMW $\hat{B}_K$ and
the exclusive $V_{cb}$ from the FNAL/MILC form factor
\cite{Bailey2014:PhysRevD.89.114504}, because their gauge ensembles
are independent.
Hence, we assume that the correlation between the FLAG $\hat{B}_K$ and
the FNAL/MILC $V_{cb}$ are negligibly small.
However, when we use the SWME $\BK$ \cite{Bae2014:prd.89.074504},
there must be an inevitable correlation with the FNAL/MILC result for
exclusive $V_{cb}$.
In this case, we consider $+50\%$ correlation and $-50\%$
anti-correlation between the SWME $\BK$ and the exclusive $V_{cb}$ to
estimate the systematic error due to the correlation between them.
The RBC/UKQCD collaboration calculated $\xi_0$ using domain-wall
fermions, which is also completely independent.
Hence, we assume that $\xi_0$ is uncorrelated with the other lattice
inputs $\BK$ and $V_{cb}$.

When we determine the value of $\epsK$, we take into account the
correlation between the SWME $\BK$ and the FNAL/MILC $V_{cb}$ and
assume that the other input parameters are uncorrelated.
We use the Monte Carlo method to calculate the $\epsK$ distribution
from the SM.
The results are cross-checked using the standard error propagation
method.

In Section \ref{sec:rev}, we review neutral kaon mixing and derive the
master formula for $\epsK$ from the SM.
Here, we give an estimate for the size of truncated
small corrections.
In Section \ref{sec:anly}, we explain each input parameter in
detail.
Here, we also explain details on how we populate input distributions
using the Monte Carlo method and how we determine errors on $\epsK$
considering different input combinations and correlations among them.
In Section \ref{sec:result}, we present the results for $\epsK$
obtained using various combinations of input parameters.
In Section \ref{sec:conclude}, we conclude.

\section{Review of $\epsK$}
\label{sec:rev}

\subsection{Effective Hamiltonian}
Let us first review the theoretical formalism of neutral kaon mixing
in the SM \cite{Buras1998:hep-ph/9806471}.
Let us consider a state that is initially (at $t=0$) a superposition
of $K^0(d\bar{s})$ and $\wbar{K}^0(s\bar{d})$:
\begin{align}
| \psi(0) \rangle &= a(0) | K^0 \rangle + b(0) | \wbar{K}^0 \rangle \,.
\end{align}
This state will evolve in time, and part of it will decay into final
states $\{f_1, f_2, \ldots\}$ as follows.
\begin{align}
| \psi(t) \rangle &= a(t) | K^0 \rangle + b(t) | \wbar{K}^0 \rangle 
\nonumber \\ & \hspace{2pc} 
+ c_1(t) | f_1 \rangle 
+ c_2(t) | f_2 \rangle + \cdots\,.
\label{eq:decay-state}
\end{align}
If we are interested in calculating only the values of $a(t)$ and
$b(t)$, but not the values of $c_i(t)$, and if the time $t$ is much
larger than the typical strong interaction scale, then we can use the
simplified formalism in Ref.~\cite{ Weisskopf1930:BF01336768,
  Weisskopf1930:BF01397406}.
In this formalism, the time evolution is described by a $2 \times 2$
effective Hamiltonian $H_\text{eff}$ that is not Hermitian, which
allows the neutral kaons to oscillate and to decay.

The neutral kaon system forms a two dimensional subspace of the
Hilbert space of the total Hamiltonian $H = H_\text{0} + H_\text{w}$.
$H_\text{0}$ is the strong interaction Hamiltonian which defines the
full Hilbert space.
Decays into different strong eigenstates are mediated by the weak
interaction Hamiltonian $H_\text{w}$, which is treated as a
perturbation.

In the 2-dimensional subspace, the time evolution of the neutral kaon
state vector can be described by the effective Hamiltonian
$H_\text{eff}$,
\begin{align}
  i\frac{d}{dt} \ket{K(t)}
  & =  H_\text{eff} \ket{K(t)} \,.
\end{align} 
The effective Hamiltonian consists of two Hermitian operators $M$ and
$\Gamma$,
\begin{equation}\label{eq:KaonMixing H_eff}
  H_\text{eff}
  = M - i\frac{\Gamma}{2} \,.
\end{equation}
The dispersive part $M$ defines masses of the neutral kaon states,
which correspond to the kaon eigenstates in nature, and the absorptive
part $\Gamma$ defines decay widths of the mass eigenstates in the
presence of the weak interaction $H_\text{w}$.
The effective Hamiltonian itself, however, is not Hermitian. 
It is a necessary consequence to take into account kaon decay
amplitudes that have final strong eigenstates which do not belong to
the neutral kaon subspace, as one can see in
Eq.~\eqref{eq:decay-state}.

The decay processes can be systematically described by the
perturbative corrections to the effective Hamiltonian of the
subspace~\cite{Wang1997:PhysRevD.55.3131}.
In the second order in $H_\text{w}$, or equivalently second order in
the Fermi coupling constant $G_F$, the results are, as shown by the
famous Wigner-Weisskopf formula~\cite{ Weisskopf1930:BF01336768,
  Weisskopf1930:BF01397406},
\begin{align}
  \label{eq:wwaM}
  M_{\alpha\beta} 
  =& m_{0}\delta_{\alpha\beta} + \mate{\alpha}{H_\text{w}}{\beta} \CL
	 &-\mathcal{P}\sum_{C} 
	  \frac{\mate{\alpha}{H_\text{w}}{C} \mate{C}{H_\text{w}}{\beta}}
	  {E_{C}-m_{K^0}} \,,\\
  \label{eq:wwaGamma}
  \Gamma_{\alpha\beta} 
  =& 2\pi \sum_{C} \mate{\alpha}{H_\text{w}}{C} 
     \mate{C}{H_\text{w}}{\beta} \delta(E_{C} - m_{0}) \,,
\end{align}
where $m_{K^0}$ is the mass of the neutral kaons $K^{0}$ and
$\wbar{K}^0$, $\mathcal{P}$ denotes the principal value, $\ket{C}$ is
an intermediate state with energy $E_C$ which belongs to the full
Hilbert space, and the summation over $C$ includes integration over
the continuous quantum numbers.
Here, we ignore a tiny experimental mass difference between $K^{0}$
and $\wbar{K}^{0}$, since we assume CPT invariance throughout this
paper.
Hence, the masses of a particle and its anti-particle are the same.

The leading correction to the off-diagonal components of
$M_{\alpha\beta}$ comes from the four-quark $\Delta S = 2$ operator of
dimension 6.
It is built from a product of two weak current-current interactions by
integrating out $W$-bosons and heavy quarks in the box loop diagrams.
This is a short distance contribution, and it is the leading effect
which is responsible for the indirect CP violation in neutral kaon
mixing.
This short distance effect is explained in Section~\ref{ssec:SD} in
detail.
If there exists a fundamental $\Delta S=2$ interaction, the so-called
superweak interaction $H_\text{sw}$, which is absent in the SM, it
also contributes to the off-diagonal components $M_{\alpha\beta}$
\cite{Winstein1993:RevModPhys.65.1113}.
Neutral kaons could decay into an intermediate state $\ket{C}$ as a
result of $\Delta S = 1$ transitions.
The parts which involve these intermediate states $\ket{C}$ in
Eq.~\eqref{eq:wwaM} and Eq.~\eqref{eq:wwaGamma} constitute the
long distance contributions.

The time independence of the effective Hamiltonian is a consequence of
the Wigner-Weisskopf approximation, which takes the interaction time
to infinity and turns the interaction adiabatically off \cite{
  Wang1997:PhysRevD.55.3131, Pokorski1987:Gauge}.
The well-known exponential decay law follows from this approximation.
So a deviation from the conventional exponential decay gives an
estimate of the accuracy of the Wigner-Weisskopf approximation.
These corrections to the exponential decay, with present and
foreseeable experimental precision \cite{ Wang1997:PhysRevD.55.3131,
  Liang2001:JPhysG.27.243}, are far beyond the precision that we pursue
here for the value of $\epsK$ in the SM.
Hence, we neglect these corrections in this paper.

Before considering explicit calculation of the matrix elements on the
right hand side of Eq.~\eqref{eq:wwaM} and Eq.~\eqref{eq:wwaGamma},
we focus on their parametrization.
From the Hermiticity of $M$ and $\Gamma$, each of them is parametrized
with 4 real variables
\begin{align}\label{eq:KaonMixing H_eff_K1K2}
  M =&
  \begin{pmatrix} 
	M_{1} & im^{\prime}+\delta_{m^\prime} \\
	-im^{\prime}+\delta_{m^\prime} & M_{2}
  \end{pmatrix} \,,\\
  \Gamma =&
  \begin{pmatrix} 
	\Gamma_{1} & i\gamma^{\prime}+\delta_{\gamma^{\prime}} \\
	-i\gamma^{\prime}+\delta_{\gamma^{\prime}} & \Gamma_{2}
  \end{pmatrix} \,.
\end{align}
Further simplification
\begin{equation}\label{eq:CPT0}
  \delta_{m^\prime} = 0 \,,\quad
  \delta_{\gamma^{\prime}} = 0
\end{equation}
follows from CPT invariance, $(CPT)O(CPT)^{-1} = O$, where $O=M,
\Gamma$, and with a specific basis made of the CP eigenstates:
$\{\ket{K_1}, \ket{K_2}\}$ \cite{Winstein1993:RevModPhys.65.1113}.

Assuming the strong interaction has CP symmetry, the neutral kaon
subspace can be spanned by the CP even $\ket{K_1}$ and odd $\ket{K_2}$
eigenstates
\begin{align}\label{eq:def of K1K2}
  \vert K_{1} \rangle 
  =& \frac{1}{\sqrt{2}} 
	\big( \ket{K^{0}} - \ket{\wbar{K^{0}}} \big) \,,\CL
  \vert K_{2} \rangle 
  =& \frac{1}{\sqrt{2}} 
	\big( \ket{K^{0}} + \ket{\wbar{K^{0}}} \big) \,.
\end{align}
We adopt a phase convention \cite{Buras1998:hep-ph/9806471}
\begin{equation}\label{eq:KaonCPConj}
  CP \ket{K^{0}} = - \ket{\wbar{K^{0}}} \,,
\end{equation}
and for time reversal $T$
\begin{equation} \label{eq:KaonTConj}
  T\ket{K^0} = - \ket{K^0} \,,\quad
  T\ket{\wbar{K^0}} = - \ket{\wbar{K^0}} \,.
\end{equation}
Here, note that the incoming state becomes an outgoing state under
time reversal and vice versa.
Then 
\begin{align}
  \ket{\wbar{K_1}} =& CPT \ket{K_1} = -\ket{K_1} \,,\CL
  \ket{\wbar{K_2}} =& CPT \ket{K_2} =  \ket{K_2} \,.
\end{align}
Then we can verify the constraints in Eq.~\eqref{eq:CPT0},
\begin{align}
  \mate{K_1}{M}{K_2}
  =& \mate{\wbar{K}_2}{(CPT)M^\dagger(CPT)^{-1}}{\wbar{K}_1} \CL
  =& -\mate{K_2}{M}{K_1} \,.
\end{align}
The same relation also holds for $\Gamma$.
Here, the Hermitian conjugate arises from the anti-unitarity of the
time reversal symmetry.

\subsection{$\epsK$ and $\teps$}
\label{ssec:mixing}
The physical states $K_S$ and $K_L$ are approximately CP even and odd,
respectively.
In other words, the physical eigenstates of the effective Hamiltonian
$H_\text{eff}$ in Eq.~\eqref{eq:KaonMixing H_eff} include a tiny
impurity ($\approx 10^{-3}$) of the opposite CP eigenstate defined in
Eq.~\eqref{eq:def of K1K2}.
The physical eigenstates can be written with small mixing parameters
$\teps_S$ and $\teps_L$,
\begin{align}
  \ket{K_{S}} =& \frac{1}{ \sqrt{ 1+\abs{\teps_S}^2 } } 
  (\ket{K_{1}} + \teps_S \ket{K_{2}}) \,,\CL
  \ket{K_{L}} =& \frac{1}{ \sqrt{ 1+\abs{\teps_L}^2 } } 
  (\ket{K_{2}} + \teps_L \ket{K_{1}}) \,.
\end{align}
Their eigenvalues are
\begin{equation}
  \lambda_{S} = \bar{\lambda} - \Delta\lambda \,,\;
  \lambda_{L} = \bar{\lambda} + \Delta\lambda \,,
\end{equation}
where
\begin{gather}
  \bar{\lambda} 
  = \frac{1}{2} \left\{ ( M_1 + M_2 )
    -\frac{i}{2} ( \Gamma_1 + \Gamma_2 ) \right\} \,,\\
  \Delta\lambda
  = \frac{1}{2} \sqrt{ 
    \Big( \Delta M + \frac{i}{2} \Delta \Gamma \Big)^2 
   + 4 \Big( m^{\prime} - \frac{i}{2}\gamma^{\prime} \Big)^2 } \,,
\end{gather}
and
\begin{equation}
  \label{eq:defDelta}
  \Delta M = M_{2} - M_{1} \,,\;
  \Delta\Gamma = \Gamma_{1}-\Gamma_{2} \,.
\end{equation}

Eliminating the eigenvalues $\lambda_{S,L}$ from the system of
eigenvalue equations
\begin{align}
  \Big( M_{1}-\frac{i}{2}\Gamma_{1}-\lambda_{S} \Big) 
  + \teps_S \Big( im^{\prime}+\frac{1}{2}\gamma^{\prime} \Big) &= 0 \,,\CL
  \teps_S \Big( M_{2}-\frac{i}{2}\Gamma_{2}-\lambda_{S} \Big) 
  - \Big( im^{\prime}+\frac{1}{2}\gamma^{\prime} \Big) &= 0 \,,\CL
  \Big( M_{2}-\frac{i}{2}\Gamma_{2}-\lambda_{L} \Big) 
  - \teps_L \Big( im^{\prime}+\frac{1}{2}\gamma^{\prime} \Big) &= 0 \,,\CL
  \teps_L \Big( M_{1}-\frac{i}{2}\Gamma_{1}-\lambda_{L} \Big) 
  + \Big( im^{\prime}+\frac{1}{2}\gamma^{\prime} \Big) &= 0
\end{align}
leads to the condition 
\begin{equation}
  ( \teps_{S,L}^{2} + 1 ) 
  \Big( im^{\prime}+\frac{1}{2}\gamma^{\prime} \Big)
  -\teps_{S,L}
  \Big( \Delta M + \frac{i}{2} \Delta \Gamma \Big) = 0
\label{eq:eps-quad}
\end{equation}
that the mixing parameters have to satisfy.
The quadratic equation in Eq.~\eqref{eq:eps-quad} has two solutions.
One of them is very small ($\approx 10^{-3}$) and the other is
very large ($\approx 10^{+3}$).
Hence, it is obvious that the two mixing parameters are equal, since
we assume that the mixing impurity is in the level of $\approx
10^{-3}$.
Hence, we will use the mixing parameter $\teps$
\begin{equation}
  \teps \equiv \teps_S = \teps_L \,.
\end{equation}

Since we know that $\mathcal{\abs{\teps}} \approx 10^{-3}$, we can
rewrite Eq.~\eqref{eq:eps-quad} as follows,
\begin{equation}
  \teps = \teps_{(0)} ( 1 + \teps^{2} ) \,,
  \label{eq:eps-iter}
\end{equation}
where
\begin{equation}
  \label{eq:LO mixing param}
  \teps_{(0)} 
  \equiv \frac{ i \Big( m^{\prime}-\frac{i}{2}\gamma^{\prime} \Big) }
  { \Delta M + \frac{i}{2} \Delta \Gamma } \,.
\end{equation}
Then, Eq.~\eqref{eq:eps-iter} can be solved iteratively near the
leading order solution $\teps_{(0)}$,
\begin{equation}
  \teps 
  = \teps_{(0)} + \teps_{(0)}^{3} 
    + 2\teps_{(0)}^{5} + 5\teps_{(0)}^{7} + \cdots \,.
\end{equation}
To complete the connection between $\teps$ and $\epsK$, we need to
consider kaon decay amplitudes
\cite{Cirigliano2011:RevModPhys.84.399}.
Define the isospin amplitude $A_{I}$ and phases $\xi_{I}$ and $\delta_{I}$
by
\begin{align}
  \label{eq:K2pipi}
  \mathcal{A}(K^0 \to \pi\pi(I)) 
	&\equiv A_I e^{i\delta_I} 
	= \abs{A_I} e^{i\xi_I} e^{i\delta_I} \,.
\end{align}
Then, in our phase convention, which is one of the most
popular conventions \cite{branco1999cp},
\begin{align}
  \label{eq:Kbar2pipi}
  \mathcal{A}(\wbar{K}^{0} \to \pi\pi(I)) 
  =& - A^\ast_I e^{i\delta_I}
	=  - \abs{A_I} e^{-i\xi_I} e^{i\delta_I} \,,
\end{align}
and
\begin{align}
  \label{eq:Keo2pipi}
  \mathcal{A}(K_1 \to \pi\pi(I)) 
  &= \sqrt{2} \Re A_I e^{i\delta_I} \,,\CL
  \mathcal{A}(K_2 \to \pi\pi(I)) 
  &= i\sqrt{2} \Im A_I e^{i\delta_I} \,,
\end{align}
where the phase $\delta_I$ is equal to the $S$-wave scattering phase shift
of the final two-pion state by the strong interaction, and the subscript $I$
represents the isospin of the final state.
Assuming isospin symmetry, this comes from Watson's theorem \cite{
  Donoghue1992:Dynamics, Cirigliano2011:RevModPhys.84.399}.
Watson's theorem is based on time reversal symmetry implicitly.
Because the final state scattering only involves $H_0$, application of
Watson's theorem concerns the time reversal symmetry of the strong
interaction.
It is equivalent to the CP symmetry, if we assume CPT invariance.
Here, note that $\xi_I$ represents the effect of the violation of
Watson's theorem.

If the weak Hamiltonian $H_\text{w}$ respected CP symmetry, which is
equivalent to time reversal symmetry under CPT invariance, then
Watson's theorem must hold to guarantee that $A_I$ must be real in
this case \cite{Donoghue1992:Dynamics}.
However, we know that $H_\text{w}$ breaks CP symmetry through the
existence of a single phase in the CKM matrix, and so it also violates
time reversal symmetry.
As a consequence, Watson's theorem is violated, and so $A_I$ becomes
complex, which generates the phase $\xi_I \ne 0$, in general.
Hence, the weak phases $\xi_I$ parametrize the direct CP violation in
the weak interaction with a non-zero phase difference, $\Im
(A_2/A_0)$, which is independent of phase convention
\cite{Buras1998:hep-ph/9806471}.
Now, let us focus on $\gamma^\prime$ and $\Delta \Gamma$ in
Eq.~\eqref{eq:LO mixing param}.
We will address $m^\prime$ and $\Delta M$ later in
Section~\ref{ssec:SD} and Section~\ref{ssec:LD}.
Let us divide both numerator and denominator of Eq.~\eqref{eq:LO
  mixing param} by $\Delta M$; we obtain
\begin{align}
  \tilde{\epsilon}_{(0)} 
  &= e^{i\theta} \sin{\theta} 
     \Big( \frac{m^{\prime}}{\Delta M} 
     - i\cot{\theta} \frac{\gamma^{\prime}}{\Delta\Gamma} \Big) \CL
  &= e^{i\theta} \sin{\theta} 
     \Big( \frac{m^{\prime}}{\Delta M} - i\xi_{0} \cot{\theta} \Big)
     + \mathcal{O}(\omega \eps^\prime)
     \CL
     & \hspace{2pc}
     + \mathcal{O}(\xi_0 \Gamma_2/\Gamma_1) \,,
  \label{eq:epsilon tilde_zero approx}
\end{align}
where
\begin{align}
  \tan{\theta} & \equiv \frac{2\Delta M}{\Delta\Gamma} \,,
  \\
  \eps^\prime
  & \equiv 
  e^{i(\delta_2-\delta_0)} \frac{i\omega}{\sqrt{2}} 
  \left( \frac{ \Im{A_2} }{ \Re{A_2} } - 
  \frac{ \Im{A_0} }{ \Re{A_0} } \right) 
  \nonumber \\
  & =   e^{i(\delta_2-\delta_0)} \frac{i\omega}{\sqrt{2}} 
  ( \xi_2 - \xi_0 ) + \mathcal{O}(\xi^3_i) \,,
  \label{eq:epsp}
\end{align}
and
\begin{align}
  \omega & \equiv \frac{ \Re{A_2} }{ \Re{A_0} } \,,\\
  \frac{ \Im{A_0} }{ \Re{A_0} } & = \tan(\xi_0) 
  = \xi_0 + \mathcal{O}(\xi^3_0) \,,
  \label{eq:xi_0} \\
  \frac{ \Im{A_2} }{ \Re{A_2} } & = \tan(\xi_2) 
  = \xi_2 +  \mathcal{O}(\xi^3_2) \,.
\end{align}
Here, we use the small angle approximation for the weak phases $\xi_0$
and $\xi_2$.

When we derive Eq.~\eqref{eq:epsilon tilde_zero approx}, we apply the
following approximation:
\begin{equation}
  \label{eq:gamma_approx}
  \frac{i\gamma^\prime}{\Delta\Gamma}
  = i\xi_0 + \mathcal{O}(\omega \eps^\prime)
    + \mathcal{O}(\xi_0 \Gamma_2/\Gamma_1) \,.
\end{equation}
It is obtained from the fact that the neutral kaon decay amplitudes
are dominated by the $I=0$ two-pion final state.
First, we can express it as follows,
\begin{align}
  \frac{i\gamma^\prime}{\Delta\Gamma}
  &= \frac{i\gamma^\prime}{\Gamma_1} \left(1 + (\Gamma_2/\Gamma_1) 
  +(\Gamma_2/\Gamma_1)^2 + \cdots \right) \,.
\end{align}
Since we know that $\Gamma_2/\Gamma_1 \approx 10^{-3}$, we can
introduce the first approximation as follows,
\begin{align}
  \frac{i\gamma^\prime}{\Delta\Gamma}
  &= \frac{i\gamma^\prime}{\Gamma_1} + \cdots \,. 
\end{align}
Using the Wigner-Weisskopf formula in Eq.~\eqref{eq:wwaGamma},
we can re-express the right-hand side as follows,
\begin{align}
  \frac{i\gamma^\prime}{\Gamma_1}
  &= \frac{ \sum_C \mate{K_1}{H_\text{w}}{C} \mate{C}{H_\text{w}}{K_2} 
     \delta(E_C - m_{K^0}) } 
     { \sum_C \mate{K_1}{H_\text{w}}{C} \mate{C}{H_\text{w}}{K_1} 
     \delta(E_C - m_{K^0}) } \,.
     \label{eq:wwa-1}
\end{align}
Here, it is obvious that the denominator is completely dominated
by the two-pion states. 
In the case of the numerator, there are contributions from two-pion
states, three-pion states, and so on.
Here, we assume that the two-pion contribution is dominant and we may
neglect the rest, which includes the (semi-)leptonic decay modes.
For example, in the case of the three-pion state, the branching ratio
between the two-pion decay and three-pion decay of $K^0_S$ is about
$3.5\times 10^{-7}$, and that for the $K^0_L$ is about 113
\cite{Agashe2014:ChinPhysC.38.090001}.
Therefore, the three-pion decay mode is suppressed by a factor of
about $6.3 \times 10^{-3}$ compared to the two-pion mode.
Similarly, we also assume that the semi-leptonic and leptonic decay
modes are so suppressed that we may neglect them in the numerator, as
in Refs.~\cite{ Winstein1993:RevModPhys.65.1113,
  Buras2010:PhysLettB.688.309}.
Therefore, as a very good approximation, we assume that the summation
in $C$ in Eq.~\eqref{eq:wwa-1} is completely dominated by the two-pion
states in both the numerator and the denominator as follows.
%
\begin{align}
  &\mathcal{A}(K_{1} \to C) \equiv \mate{C}{H_\text{w}}{K_1}
  \nonumber \\
  & = \delta_{C, \pi\pi(0)} \mate{\pi\pi(0)}{H_\text{w}}{K_1}
  + \delta_{C, \pi\pi(2)} \mate{\pi\pi(2)}{H_\text{w}}{K_1}
  \nonumber \\
  & \hspace{2pc} + \cdots
  \nonumber \\
  & = \delta_{C, \pi\pi(0)} \sqrt{2} (\Re A_0) e^{i\delta_0}
  + \delta_{C, \pi\pi(2)} \sqrt{2} (\Re A_2) e^{i\delta_2}
  \nonumber \\
  & \hspace{2pc} + \cdots \,.
  \label{eq:K1->C}
\end{align}
Similarly,
\begin{align}
  &\mathcal{A}(K_{2} \to C) \equiv \mate{C}{H_\text{w}}{K_2}
  \nonumber \\
  & = \delta_{C, \pi\pi(0)} \mate{\pi\pi(0)}{H_\text{w}}{K_2}
  + \delta_{C, \pi\pi(2)} \mate{\pi\pi(2)}{H_\text{w}}{K_2}
  \nonumber \\
  & \hspace{2pc} + \cdots
  \nonumber \\
  & = \delta_{C, \pi\pi(0)} i\sqrt{2} (\Im A_0) e^{i\delta_0}
  + \delta_{C, \pi\pi(2)} i\sqrt{2} (\Im A_2) e^{i\delta_2}
  \nonumber \\
  & \hspace{2pc} + \cdots
  \label{eq:K2->C}
\end{align}

Using Eqs.~\eqref{eq:K1->C}, \eqref{eq:K2->C}, and
\eqref{eq:wwa-1}, we can obtain the following result:
\begin{align}
  \frac{i\gamma^\prime}{\Gamma_1}
  &= \frac{ 2 i (\Re A_0) (\Im A_0) + 2 i (\Re A_2)(\Im A_2) + \cdots }
   { 2 (\Re A_0)^2  + 2 (\Re A_2)^2 + \cdots } \CL
  &= i \left[ \frac{\Im A_0}{\Re A_0}
   + \left(\frac{\Re A_2}{\Re A_0} \right)^2 
   \left\{\frac{\Im A_2}{\Re A_2} - \frac{\Im A_0}{\Re A_0}\right\}
   + \cdots \right] \CL
  &= i\xi_0 + \sqrt{2} \omega \eps^\prime e^{i(\delta_0 - \delta_2)} 
   + \cdots \,.
  \label{eq:expand_Gam2overGam1}
\end{align}
Here, we know that $\xi_0 \approx 10^{-4}$ and 
$\sqrt{2} \omega \eps^\prime \approx 10^{-7}$.
Hence we may safely neglect the $\omega \eps^\prime$ term in
Eq.~\eqref{eq:expand_Gam2overGam1} within the precision that we pursue
in this paper.
This leads to the approximation in Eq.~\eqref{eq:gamma_approx}.

In terms of isospin amplitudes, $\epsK$ in Eq.~\eqref{eq:epsK_def} can
be written
\begin{align}
  \epsK 
  &= \frac{ i\Im{A_{0}} + \teps \Re{A_{0}} }
          { \Re{A_{0}} + i\teps \Im{A_{0}} }
   = \frac{\teps + i \xi_0}{1 + i\teps \xi_0} \CL
  &= ( \teps + i\xi_{0} )
     ( 1 - i\teps \xi_0 + \cdots ) \CL
  &= \teps + i\xi_{0} - i \teps^2 \xi_0 + \teps \xi_{0}^2 + \cdots \CL
  &= \teps_{(0)} + i\xi_{0} + \teps_{(0)}^3 - i \teps^2_{(0)} \xi_0 
   + \cdots \CL
	\label{eq:epsilon-lo}
  &= \teps_{(0)} + i \xi_{0} 
     + \mathcal{O}( {\teps_{(0)}}^{3} )\,.
\end{align}

Finally, using Eq.~\eqref{eq:epsilon tilde_zero approx} and
Eq.~\eqref{eq:epsilon-lo}, we obtain
\begin{align}
  \label{eq:epsK}
  \epsK
  &= e^{i\theta} \sin{\theta} \Big( \frac{ m^{\prime} }{ \Delta M } 
  + \xi_{0} \Big)
   + \mathcal{O}(\omega \eps^\prime)
   \CL
  & \hspace{2pc}
   + \mathcal{O}(\xi_0 \Gamma_2/\Gamma_1) \,.
\end{align}
Here, we keep only the first two terms from Eq.~\eqref{eq:epsilon-lo}.
The size of those corrections that we neglect in this paper is much
smaller than the experimental precision of $\epsK$, as one can see
in Eq.~\eqref{eq:epsK_exp}.

\subsection{Short Distance Contribution}
\label{ssec:SD}

The matrix element $m^{\prime}$ can be calculated from the
Wigner-Weisskopf formula given in Eq.~\eqref{eq:wwaM}.
A short distance contribution $m^\prime_\text{SD}$ to $m^\prime$ is 
\begin{align}\label{eq:mSD}
  2m_{K^{0}} \cdot im^{\prime}_\text{SD} 
  &= \mate{K_{1}}{\mathcal{H}_\text{SD}^{(6)}}{K_{2}} \CL
  &= \frac{1}{2} ( 
  \mate{K^{0}}{\mathcal{H}_\text{SD}^{(6)}}{\wbar{K^{0}}}
    - \mate{\wbar{K^{0}}}{\mathcal{H}_\text{SD}^{(6)}}{K^{0}}
  ) \CL
  &= \frac{1}{2} (
    \mate{\wbar{K^{0}}}{\mathcal{H}_\text{SD}^{(6)}}{K^{0}}^{\ast}
    - \mate{\wbar{K^{0}}}{\mathcal{H}_\text{SD}^{(6)}}{K^{0}}
  ) \CL
  &= -i \Im \mate{\wbar{K^{0}}}{\mathcal{H}_\text{SD}^{(6)}}{K^{0}} \,.
  \end{align} 
The factor $2m_{K^0}$ comes from the normalization condition for the
external kaon states.

\begin{figure}[!t]
\includegraphics[width=\columnwidth]{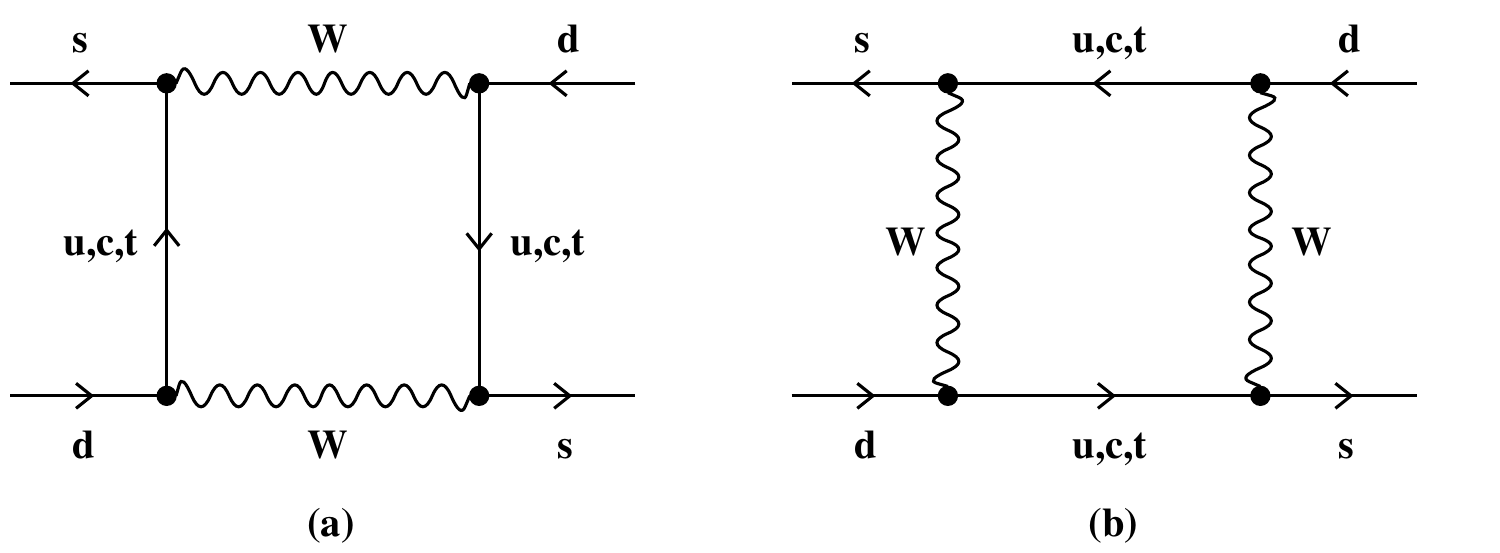}
\caption{Box diagrams for the $K^0-\wbar{K}^0$ mixing.}
\label{fig:box}
\end{figure}

In the SM, the Hamiltonian density $\mathcal{H}_\text{SD}^{(6)}$
represents the leading short distance term in the $\Delta S = 2$
effective weak Hamiltonian, which is constructed from the box
diagrams.
For a scale below charm quark threshold $\mu < \mu_{c} \approx
\mathcal{O}(m_{c})$,
\begin{align}
  \label{eq:SM Delta S=2 effective H} 
  \mathcal{H}_\text{SD}^{(6)} 
  &= \frac{G_F^2}{16\pi^2} M_W^2
     [\lambda_c^2 \eta_{cc} S_0(x_c) + \lambda_t^2 \eta_{tt} S_0(x_t) \CL
  & \quad + 2\lambda_c \lambda_t \eta_{ct} S_0(x_c,x_t)] 
    b(\mu) O_{LL}^{\Delta S = 2}(\mu) + h. c. \;.
\end{align}
Here, the dimension-6 local four fermion operator which comes from the
well-known box diagrams in Fig.~\ref{fig:box} is
\begin{equation}
  O_{LL}^{\Delta S = 2}(\mu) 
  \equiv \bar{s}\gamma_{\mu}(1-\gamma_5)d 
         \bar{s}\gamma^{\mu}(1-\gamma_5)d \,.
\end{equation}
By integrating out the heavy degrees of freedom in the loops of the
box diagrams, we obtain the Inami-Lim functions
\cite{Inami1980:ProgTheorPhys.65.297} as follows,
\begin{align}
  \label{eq:InamiFn}
  S_{0}(x_i)
  &= x_i \bigg[ \frac{1}{4} + \frac{9}{4(1-x_i)} - \frac{3}{2(1-x_i)^2}
    - \frac{3x_i^2 \ln x_i}{(1-x_i)^3}  \bigg] \,,\CL
  S_{0}(x_i,x_j) 
  &= \Bigg\{ \frac{x_i x_j}{x_i-x_j} 
     \bigg[ \frac{1}{4} + \frac{3}{2(1-x_i)}
    - \frac{3}{4(1-x_i)^2} \bigg] \ln x_i \CL
  &\quad - ( i \leftrightarrow j) \Bigg\} - \frac{3x_i x_j}{4(1-x_i)(1-x_j)} \,,
\end{align}
where, $i = c, t$, $x_i = m_i^2/M_W^2$, $m_i = m_i(m_i)$ is the scale
invariant $\overline{\text{MS}}$ quark mass \cite{
  Chetyrkin2000:CompPhysComm.133.43}, and $M_W$ is the $W$-boson pole
mass.
The $u$-quark contribution is rearranged into $c$ and $t$ terms by
imposing a unitarity condition,
\begin{align}
  \label{eq:ut_cond}
  & \lambda_u + \lambda_c + \lambda_t = 0 \,,
  \\
  & \lambda_{i} \equiv V_{is}^{\ast} V_{id} \,,
  \nonumber
\end{align}
and then the effective Hamiltonian $\mathcal{H}_\text{SD}^{(6)}$ is
re-expressed with $c$ and $t$ terms.
In Eq.~\eqref{eq:InamiFn}, an approximation $m_u^2/M_W^2 = 0$ is used.
Each pair of vertices for $W$-boson interchange gives the products of
the CKM matrix elements $\lambda_{i} = V_{is}^{\ast} V_{id}$.

Besides a zeroth order $\alpha_s^0$ QCD effect dealt with by the
Inami-Lim functions $S_0$, $\eta_{ij}$ with $i,j=c,t$ incorporate QCD
corrections of higher order in $\alpha_s$.
These are obtained by resumming large logarithms with the
renormalization group evolution \cite{Herrlich1996:NuclPhysB.476.27}.
To make it scale and renormalization scheme independent, the
renormalization group running factor with 3-flavors $b(\mu)$ is
factored out,
\begin{align}
  b(\mu) &= [ \alpha_{s}^{(3)}(\mu) ]^{-2/9} K_+(\mu) \,,
\end{align}
where $K_+(\mu)$ is given in Eq.~\eqref{eq:rg_K_inv} 
of Appendix \ref{app:eta_cc}.

It is combined with the hadronic matrix elements of the four fermion
operator $\mathcal{O}_{LL}^{\Delta S=2}(\mu)$ and used to define a
renormalization group invariant quantity $\BK$,
\begin{equation}
  \BK \equiv B_K(\mu) b(\mu) \,,
\end{equation}
where
\begin{align}
  B_K(\mu) 
  &\equiv
  \frac{ \mate{ \bar{K}^{0} }{ O_{LL}^{\Delta S = 2}(\mu) }{ K^{0} } }
       { \frac{8}{3} 
         \mate{\bar{K}^0}{\bar{s} \gamma_{\mu}\gamma_5 d}{0} 
         \mate{0}{\bar{s} \gamma^{\mu}\gamma_5 d}{K^0} } \CL
  &= \frac{ \mate{\bar{K}^0}{O_{LL}^{\Delta S = 2}(\mu)}{K^0} }
          { \frac{8}{3} F_K^2 m_{K^0}^2 }
\end{align}
can be calculated from lattice QCD at a common scale such as $\mu =
2\;\GeV$.
$F_K$ is the kaon decay constant.

Inserting Eq.~\eqref{eq:SM Delta S=2 effective H} into
Eq.~\eqref{eq:mSD}, we can identify the short distance
contribution to $m^\prime$ as follows,
\begin{equation}
  \label{eq:mprimeSD}
  m^\prime_\text{SD}
  = \frac{G_F^2}{6\pi^2} F_K^2 m_{K^0} M_W^2 \BK X_\text{SD} \,,
\end{equation}
where
\begin{align}
  X_\text{SD} &= \Im\lambda_t \Big[ \Re\lambda_c \eta_{cc} S_0(x_c)
    -\Re\lambda_t \eta_{tt} S_0(x_t) \CL & \quad - (\Re\lambda_c -
    \Re\lambda_t) \eta_{ct} S_0(x_c,x_t) \Big] \,.
\end{align}
Here, we use another unitarity identity, $\Im\lambda_t =
-\Im\lambda_c$.
It can be shown from the unitarity condition of Eq.~\eqref{eq:ut_cond}
and noting that $\lambda_u$ is real in the standard parametrization.

With the Wolfenstein parametrization for the CKM matrix elements
\cite{Buras1998:hep-ph/9806471},
\begin{align}
  \label{eq:rec}
  \Re\lambda_c 
  &= -\lambda \left(1-\frac{\lambda^2}{2}\right)
     \Bigg[ 1 - \frac{\lambda^4}{8} - A^2 \lambda^4 (1-\bar{\rho}) \Bigg]
     \,,\\
  \label{eq:ret}
  \Re\lambda_t 
  &= -\left(1-\frac{\lambda^2}{2}\right) A^2 {\lambda}^5 (1-\bar{\rho}) \,,\\
  \label{eq:imt}
  \Im\lambda_t 
  &= \eta A^2 {\lambda}^5 \,,
\end{align}
where
\begin{equation}
  \bar{\rho} = \rho \left(1-\frac{\lambda^{2}}{2}\right) \,,\quad 
  \bar{\eta} = \eta \left(1-\frac{\lambda^{2}}{2}\right) \,.
\end{equation}
They are accurate to $\mathcal{O}(\lambda^5)$.
Here, we have neglected terms of $\mathcal{O}(\lambda^7)$.
Then
\begin{align}
  \label{eq:SDfactorX}
  X_\text{SD} &= \bar{\eta}\lambda^2 \abs{V_{cb}}^2 
       \Bigg[ \abs{V_{cb}}^2 (1-\bar{\rho}) \eta_{tt} S_0(x_t) (1 + r) \CL
    &  \quad + \left(1-\frac{\lambda^4}{8}\right) 
       \left\{ \eta_{ct} S_0(x_c,x_t) - \eta_{cc} S_0(x_c) \right\} 
       \Bigg] \,,
\end{align}
where $\displaystyle r = \{\eta_{cc} S_0(x_c) - 2\eta_{ct}
S_0(x_c,x_t)\}/\{\eta_{tt} S_0(x_t)\}$. 
Here, note that we replace $A$ by $V_{cb}$, using the relation
$\displaystyle \abs{V_{cb}} = A \lambda^{2} +
\mathcal{O}(\lambda^{8})$.

\subsection{Long Distance Contribution}
\label{ssec:LD}
In the previous section, Section~\ref{ssec:SD}, we explain the
short distance contribution of the effective Hamiltonian
$\mathcal{H}_\text{SD}^{(6)}$ to $m^\prime$.
Here, we would like to address the effect of the long distance
contribution to $m^\prime$.

The parts of second order in $H_\text{w}$ in Eq.~\eqref{eq:wwaM} and
Eq.~\eqref{eq:wwaGamma} correspond to the long distance contributions.
The long distance contribution $m^\prime_\text{LD}$ of $m^\prime$ is
\begin{align}
  \label{eq:mLD}
  m^\prime_\text{LD}
  &= -i \mathcal{P}\sum_{C} 
	  \frac{\mate{K_1}{H_\text{w}}{C} \mate{C}{H_\text{w}}{K_2}}
    {m_{K^0}-E_{C}} \CL
  &= -\Im \left[ \mathcal{P}\sum_{C} 
    \frac{\mate{\wbar{K}^0}{H_\text{w}}{C} \mate{C}{H_\text{w}}{K^0}}
    {m_{K^0}-E_{C}}  \right] + \delta m^\prime_\text{LD} \,, \CL
  \delta m^\prime_\text{LD} &= \frac{1}{2}\mathcal{P}\sum_{C}
  \frac{ \abs{\mate{K_0}{H_\text{w}}{C}}^2 - 
    \abs{\mate{\wbar{K}^0}{H_\text{w}}{C}}^2 }{m_{K^0}-E_{C}}
    \,.
\end{align}
Here, note that $\delta m^\prime_\text{LD}$ vanishes due to CPT
invariance,
\begin{align}
\delta m^\prime_\text{LD} = 0\,.
\end{align}
The absorptive part $\gamma^\prime$, which comes entirely from the
long distance effect, is treated in the previous section,
Section~\ref{ssec:mixing}.

The net contribution $\xi_\text{LD}$ to $\epsK$ in
Eq.~\eqref{eq:epsK_SM_1}, which comes from $m^\prime_\text{LD}$, was
estimated to be the same order of magnitude as $\xi_0$ using chiral
perturbation theory \cite{Buras2010:PhysLettB.688.309}.
They claim that $\xi_\text{LD} = -0.4(3) \xi_0$ and that
$\xi_\text{LD}$ is at most a $4\%$ correction to $\epsK$.
This claim is consistent with the estimate of about $2\%$ in
Ref.~\cite{ Christ2012:PhysRevD.88.014508}.

Following the estimate of the long distance contribution
$m^\prime_\text{LD}$, it was claimed in Ref.~\cite{
  Buras2010:PhysLettB.688.309} that this contribution should be
incorporated.
Here, we treat the long distance effect of $m^\prime_\text{LD}$
as a systematic error in the error budget of $\epsK$. 

The theoretical expression for the mass difference $\Delta M$ defined
by Eq.~\eqref{eq:defDelta} is 
\begin{equation}
  \Delta M =
  2\Re \mate{\wbar{K^{0}}}{H_\text{SD}^{(6)}}{K^{0}} + \Delta M_{\text{LD}} \,,
\end{equation}
\begin{equation}
  \Delta M_\text{LD}
  = 2 \Re \left[ \mathcal{P}\sum_{C} 
    \frac{\mate{\wbar{K}^0}{H_\text{w}}{C} \mate{C}{H_\text{w}}{K^0}}
    {m_{K^0}-E_{C}}  \right] \,.
\end{equation}
There has been an attempt to calculate $\Delta M_\text{LD}$ in lattice
QCD \cite{ Bai:2014cva, Christ2012:PhysRevD.88.014508}.
Since the precision of lattice results is not as good as that of
experiment, we use the experimental results for $\Delta M_K$ in this
paper.

Hence, we take the experimental value of $\Delta M_K$ for $\Delta M$
in Eq.~\eqref{eq:epsK}.
This is a very good approximation,
\begin{align}
  \label{eq:DeltaMK}
  \Delta M_{K} &= M_{L} - M_{S} = \mathrm{Re}(\lambda_{L}-\lambda_{S}) \nonumber\\ 
  &= \mathrm{Re} \sqrt{ \Big( \Delta M + \frac{i}{2} \Delta\Gamma \Big)^{2} ( 1 - 4{\tilde{\epsilon}_{(0)}}^{2}) } \nonumber\\
  &= \Delta M \cdot \mathrm{Re} \Big[ ( 1 + i\cot{\theta} ) \sqrt{ 1 - 4{\tilde{\epsilon}_{(0)}}^{2} } \Big] \nonumber\\
  &= \Delta M \Big( 1 - 2 \Re \left[ \tilde{\epsilon}_{(0)}^{2} 
  (1 + i \cot\theta) \right] + \mathcal{O}( \tilde{\epsilon}_{(0)}^{4}) 
  \Big) \,.
\end{align}
Here, note that $\theta \cong \pi/4$ and $\teps_{(0)} \cong \epsK$.
Hence, the difference between $\Delta M_K$ and $\Delta M$ is of
$\mathcal{O}(\Delta M \teps_{(0)}^2)$.
This small correction can make a change of
$\mathcal{O}(\teps_{(0)}^3)$ in $\epsK$.
Here, note that $\mathcal{O}(\teps_{(0)}^3) \ll \mathcal{O}(\omega
\eps^\prime)$.
Hence, this is so small that we neglect it.

\subsection{Master Formula: $\epsK$}
From Eqs.~\eqref{eq:epsK}, \eqref{eq:mprimeSD}, \eqref{eq:SDfactorX},
and \eqref{eq:mLD}, the phenomenological expression for the indirect
CP violation parameter in the SM is
\begin{align}
  \label{eq:epsK_SM_0}
  \epsK
  =& e^{i\theta} \sqrt{2}\sin{\theta} 
  \Big( C_{\eps} \hat{B}_{K} X_\text{SD}
  + \frac{ \xi_{0} }{ \sqrt{2} } + \xi_\text{LD} \Big) \CL
   &+ \mathcal{O}(\omega \eps^\prime)
   + \mathcal{O}(\xi_0 \Gamma_2/\Gamma_1) \,,
\end{align}
where
\begin{align}
  C_{\eps} 
	&= \frac{ G_{F}^{2} F_K^{2} m_{K^{0}} M_{W}^{2} }
		   { 6\sqrt{2} \pi^{2} \Delta M_{K} } \,,
\label{eq:C_eps}
\\
  \xi_\text{LD} &=  \frac{m^\prime_\text{LD}}{\sqrt{2} \Delta M_K} \,.
\label{eq:xiLD}
\end{align}
Here, $\xi_\text{LD}$ is the long distance effect of $\approx 2\%$
\cite{ Christ2012:PhysRevD.88.014508}.
Precise theoretical evaluation of $\xi_\text{LD}$ from lattice QCD is
not available yet.
Hence, we do not include this effect in the central value of
$\epsK$, but we take it as a systematic error in the error
budget of $\epsK$.

The correction terms $\mathcal{O}(\omega \eps^\prime)$ and
$\mathcal{O}(\xi_0 \Gamma_2/\Gamma_1)$ are of order $10^{-7}$, and we
also neglect them in this analysis.

In Eq.~\eqref{eq:SDfactorX}, the parameter $r$ is very small ($\approx
10^{-4}$) and also $\lambda^4/8 \approx 10^{-4}$.
Hence, if we neglect these small terms in Eq.~\eqref{eq:SDfactorX}, we
can obtain the same formula as in
Ref.~\cite{Buras2008:PhysRevD.78.033005}.
However, in this paper we keep both the $r$ parameter and the
$\lambda^4/8$ term in Eq.~\eqref{eq:SDfactorX}, even though they make
no difference to our conclusion.

In Ref.~\cite{Buras2008:PhysRevD.78.033005}, the multiplicative factor
$\kappa_{\eps}$ was introduced to incorporate long distance effects
$\xi_\text{LD}$, the small additive correction $\xi_{0}$, and
deviation of the angle $\theta$ from the value $45^{\circ}$.
Since $\xi_0$ can be estimated from lattice QCD \cite{
  Blum2011:PhysRevLett.108.141601}, we can treat this small
contribution to $\epsK$ explicitly.

\section{Data analysis}
\label{sec:anly}

\subsection{Input Parameters}
\label{ssec:iparam}
%
%
The CKMfitter and UTfit groups provide the Wolfenstein parameters
$\lambda, \bar{\rho}, \bar{\eta}$ and $A$ from the global UT fit.
Here, we use $\lambda, \bar{\rho}, \bar{\eta}$ from CKMfitter \cite{
  Charles:2004jd, Hocker:2001xe} and UTfit \cite{ Bona:2005vz,
  Bona:2007vi}, and we use $V_{cb}$ instead of $A$,
Eq.~\eqref{eq:SDfactorX}.
The parameters $\lambda$, $\bar{\rho}$, and $\bar{\eta}$ are
summarized in Table~\ref{tbl:in-wolf}.

The parameters $\epsK, \hat{B}_K$, and $V_{cb}$ are inputs to the
global UT fit.
Hence, the Wolfenstein parameters extracted from the global UT fit of
the CKMfitter and UTfit groups contain unwanted dependence on the
$\epsK$ calculated from the master formula,
Eq.~\eqref{eq:epsK_SM_0}.
To self-consistently determine $\epsK$, we take another input set from
the angle-only fit (AOF) in Ref.~\cite{Bevan2013:npps241.89}.
The AOF does not use $\epsK, \hat{B}_K$, and $V_{cb}$ as inputs
to determine the UT apex of $\bar{\rho}$ and $\bar{\eta}$
\cite{Bevan2013:npps241.89}.
The AOF gives the UT apex $(\bar{\rho}, \bar{\eta})$ but not
$\lambda$.
We can take $\lambda$ independently from the CKM matrix element
$V_{us}$, because this is parametrized by
\begin{equation}
  \abs{V_{us}} = \lambda + \mathcal{O}(\lambda^7) \,.
\end{equation}
Here we use the average of results extracted from the $K_{\ell3}$ and
$K_{\mu2}$ decays \cite{Agashe2014:ChinPhysC.38.090001}.  
\begin{table}[!th]
  \caption{Wolfenstein Parameters}
  \label{tbl:in-wolf}
  \renewcommand{\arraystretch}{1.2}
  \begin{ruledtabular}
  \begin{tabular}{cccc}
  & CKMfitter & UTfit & AOF \cite{Bevan2013:npps241.89} \\ \hline
  $\lambda$
  & $0.22537(61)$
  /\cite{Agashe2014:ChinPhysC.38.090001}
  & $0.2255(6)$ 
  /\cite{Agashe2014:ChinPhysC.38.090001}
  & $0.2253(8)$
  /\cite{Agashe2014:ChinPhysC.38.090001}
  \\ \hline
  $\bar{\rho}$
  & $0.117(21)$ 
  /\cite{Agashe2014:ChinPhysC.38.090001}
  & $0.124(24)$ 
  /\cite{Agashe2014:ChinPhysC.38.090001}
  & $0.139(29)$
  /\cite{UTfit2014PostMoriondSM:web} 
  \\ \hline
  $\bar{\eta}$ 
  & $0.353(13)$ 
  /\cite{Agashe2014:ChinPhysC.38.090001}
  & $0.354(15)$ 
  /\cite{Agashe2014:ChinPhysC.38.090001}
  & $0.337(16)$
  /\cite{UTfit2014PostMoriondSM:web}
  \end{tabular}
  \end{ruledtabular}
\end{table}

The input values that we use for $V_{cb}$ are summarized in
Table~\ref{tbl:in-Vcb}. The inclusive determination considers the
following inclusive decays: $ B \to X_c l \nu$ and $B \to X_s
\gamma $.
Moments of lepton energy, hadron masses, and photon energy are
measured from the relevant decay.
Those moments are fit to theoretical expressions which are obtained by
applying the operator product expansion (OPE) to the decay amplitude
with respect to the strong coupling $\alpha_s$ and inverse heavy quark
mass $\Lambda/m_b$.
There are two schemes for the choice of $b$ quark mass $m_b$ in the
heavy quark expansion: the kinetic scheme and the 1S scheme \cite{
  Agashe2014:ChinPhysC.38.090001, Alberti2014:PhysRevLett.114.061802}.
We use the value obtained using the kinetic scheme \cite{
  Alberti2014:PhysRevLett.114.061802}, which has somewhat larger
errors and also was updated more recently.\footnote{ In
  Ref.~\cite{Alberti2014:PhysRevLett.114.061802}, inclusive $V_{cb}$
  is determined using the semi-leptonic $B$ decays but not radiative
  $B$ decays. }

The exclusive determination considers the semi-leptonic decay of
$\bar{B}$ to $D$ or $D^{\ast}$.
Here, we use the most up-to-date value from the FNAL/MILC lattice
calculation of the form factor $\mathcal{F}(w)$ of the semi-leptonic
decay $\bar{B}\to D^{\ast}\ell\bar{\nu}$ at zero-recoil ($w=1$)
\cite{Bailey2014:PhysRevD.89.114504}.
The authors of Ref.~\cite{ Bailey2014:PhysRevD.89.114504} used the
Wilson clover action for the heavy quarks, which is tuned by the
Fermilab interpretation \cite{ El-khadra:PhysRevD.55.3933} via heavy
quark effective theory \cite{ Harada:2001fj, Harada:2001fi,
  Kronfeld:2000ck}, with the MILC $N_f=2+1$ asqtad gauge ensembles
\cite{ Bazavov:2009bb}.
The heavy quark symmetry and heavy quark effective theory play a key
role throughout their strategies.
Considering about a $1\%$ enhancement by the electromagnetic
correction $\abs{\bar{\eta}_\text{EM}}$, they combined their lattice
result with the HFAG average \cite{Amhis2012:HFAG} of experimental
values $\mathcal{F}(1) \abs{\bar{\eta}_\text{EM}} \abs{V_{cb}}$ to
extract $\abs{V_{cb}}$.
%
%
\begin{table}[!th]
  \caption{ Inclusive and exclusive $\abs{V_{cb}}$ in units of
    $10^{-3}$. Here, Kin.~represents the kinetic scheme in the heavy
    quark expansion, and 1S, the 1S scheme.}
  \label{tbl:in-Vcb}
  \renewcommand{\arraystretch}{1.2}
  \begin{ruledtabular}
  \begin{tabular}{ccc}
  Inclusive (Kin.) & Inclusive (1S) & Exclusive \\ \hline
  $42.21(78)$
  /\cite{Alberti2014:PhysRevLett.114.061802}
  & $41.96(45)(07)$ 
  /\cite{Bauer2004:PhysRevD.70.094017}
  & $39.04(49)(53)(19)$
  /\cite{Bailey2014:PhysRevD.89.114504}
  \end{tabular}
  \end{ruledtabular}
\end{table}

There has been significant progress in unquenched QCD studies in
lattice gauge theory since 2000.
This progress makes several lattice calculations of $\BK$ available at
$N_f=2+1$ \cite{ Bae2012:PhysRevLett.109.041601,
  Aoki2011:PhysRevD.84.014503, Aubin2010:PhysRevD.81.014507,
  Durr2011:PhysLettB.705.477}.
FLAG provides various lattice results for $\BK$ with $N_f=2+1$ and the
lattice average \cite{ Aoki2013:hep-lat.1310.8555}.
Here, we use the $N_f=2+1$ FLAG average in Ref.~\cite{
  Aoki2013:hep-lat.1310.8555} and the SWME result as inputs, which are
summarized in Table~\ref{tbl:in-BK}.
FLAG uses the SWME result from Ref.~\cite{
  Bae2012:PhysRevLett.109.041601}, which is not much different from
the most up-to-date value \cite{Bae2014:prd.89.074504} that we use in
this analysis.
The BMW calculation \cite{ Durr2011:PhysLettB.705.477} quotes the
smallest error, and it dominates the FLAG average.
The SWME result \cite{Bae2014:prd.89.074504} quotes a larger error,
and its value deviates most from the FLAG average.
\begin{table}[!th]
  \caption{$\BK$}
  \label{tbl:in-BK}
  \renewcommand{\arraystretch}{1.2}
  \begin{ruledtabular}
  \begin{tabular}{cccc}
  & FLAG & SWME & \\ \hline
  & $0.7661(99)$
	/\cite{Aoki2013:hep-lat.1310.8555} 
	& $0.7379(47)(365)$
	/\cite{Bae2014:prd.89.074504}
  &
	\end{tabular}
  \end{ruledtabular}
\end{table}

%
The RBC/UKQCD collaboration provides lattice results for $\mathrm{Im}
A_2$ and $\xi_0$ \cite{ Blum2011:PhysRevLett.108.141601}.
They obtain $\xi_0$ (defined in Eq.~\eqref{eq:xi_0}) using the
relation
\begin{equation}
  \label{eq:xi0-from-epsprime}
  \mathrm{Re} \Big( \frac{\eps^\prime}{\epsK} \Big) 
  =	\frac{\cos(\phi_{\eps^\prime}-\phi_\eps)}
         {\sqrt{2}\abs{\epsK}}  
		\frac{\mathrm{Re}A_{2}}{\mathrm{Re}A_{0}} 
		\Big( \frac{\mathrm{Im}A_{2}}{\mathrm{Re}A_{2}} - \xi_{0} \Big) \,.
\end{equation}
In this relation, they use the lattice result for $\Im A_2$ and take
the experimental values for the remaining parameters to obtain
$\xi_0$.
In particular, they use the experimental value of $\epsK$ as an input
parameter to determine $\xi_0$.
However, the error is dominated by the experimental error of
$\mathrm{Re}(\eps^\prime/\epsK)$, which is $\approx 14\%$.
In the numerator, $\cos(\phi_{\varepsilon^\prime} - \phi_\varepsilon)$
is approximated by $1$, because the two phases are very close to each
other \cite{Agashe2014:ChinPhysC.38.090001},
\begin{align}
  \phi_{\eps} &= 43.52(5) \,, \\
  \phi_{\eps^\prime} &= 42.3(15) \,.
\end{align}
The final result for $\xi_0$ in Ref.~\cite{
  Blum2011:PhysRevLett.108.141601} is
\begin{align}
\xi_0 &= -1.63(19)(20) \times 10^{-4} \,.
\label{eq:xi_0:rbc}
\end{align}

The magnitude of $\xi_\text{LD}$ is about 1.6\%~\cite{
  Christ2012:PhysRevD.88.014508}.  We incorporate the systematic
uncertainty in $\varepsilon_K$ due to neglecting $\xi_\text{LD}$ by
treating $\xi_\text{LD}$ as a Gaussian distribution about zero with a
width of 1.6\%,
\begin{align}
\xi_\text{LD} &= (0 \pm 1.6) \% \,.
\label{eq:xi_LD:2}
\end{align}
%

%
%
The factor $\eta_{tt}$ is given at next-to-leading order (NLO) in
Ref.~\cite{Buras2008:PhysRevD.78.033005}.
Other factors $\eta_{ct}$ and $\eta_{cc}$ are given at
next-to-next-to-leading order (NNLO) in
Refs.~\cite{Brod2010:prd.82.094026} and
\cite{Brod2011:PhysRevLett.108.121801}, respectively.
The NNLO values of $\eta_{ct}$ and $\eta_{cc}$ are larger than the NLO
results in Ref.~\cite{ Buras2008:PhysRevD.78.033005}:
\begin{align}
\eta_{ct}^\text{NLO} &= 0.47(4)\,,
\\
\eta_{ct}^\text{NNLO} &= 0.496(47)\,,
\\
\eta_{cc}^\text{NLO} &= 1.43(23) \,,
\\
\eta_{cc}^\text{NNLO} &= 1.72(27) \,.
\end{align}
Here, we quote the NNLO result for $\eta_{cc}$ from SWME in Table
\ref{tbl:cmp-etacc}, which is a major update to our previous analysis
\cite{ Bailey:2014qda}.
In the case of $\eta_{cc}$, the NNLO correction is as large as the NLO
correction.
Hence, the convergence of the perturbative series in $\eta_{cc}$ is in
question \cite{ Brod2011:PhysRevLett.108.121801}.

In Ref.~\cite{ Buras2013:EurPhysJC.73.2560}, they claim that the error
is overestimated for the NNLO value of $\eta_{cc}$ given in
Ref.~\cite{ Brod2011:PhysRevLett.108.121801}.
Hence, in order to check the claim, we follow the renormalization
group (RG) evolution for $\eta_{cc}$ described in Ref.~\cite{
  Brod2011:PhysRevLett.108.121801} to produce the NNLO value of
$\eta_{cc}$.
The results are summarized in Table \ref{tbl:cmp-etacc}.
In this table, note that the results are consistent with
one another within the systematic errors.
Here, ``SWME'' represents our evaluation of $\eta_{cc}$, which is
essentially identical to that of Ref.~\cite{
  Buras2013:EurPhysJC.73.2560}.
Details of our results are explained in Appendix \ref{app:eta_cc}.
In this paper, we use the SWME result for $\eta_{cc}$ to obtain
$\epsK$. 
\begin{table}[!th]
  \caption{Results of $\eta_{cc}$ at NNLO.}
  \label{tbl:cmp-etacc}
  \renewcommand{\arraystretch}{1.2}
  \begin{ruledtabular}
  \begin{tabular}{ccc}
  collaboration & Value & Ref. \\ \hline\hline
  Brod and Gorbahn & $1.86(76)$
  &\cite{Brod2011:PhysRevLett.108.121801} \\ \hline
  Buras and Girrbach & $1.70(21)$ 
  &\cite{Buras2013:EurPhysJC.73.2560} \\ \hline
  SWME & $1.72(27)$
  &Appendix \ref{app:eta_cc}
  \end{tabular}
  \end{ruledtabular}
\end{table}

The input values for $\eta_{ij}$ that we use in this paper are
summarized in Table~\ref{tbl:in-qcdeta}.
\begin{table}[!th]
  \caption{QCD corrections}
  \label{tbl:in-qcdeta}
  \renewcommand{\arraystretch}{1.2}
  \begin{ruledtabular}
  \begin{tabular}{clc}
  Input & Value & Ref. \\ \hline\hline
  $\eta_{cc}$ & $1.72(27)$
  &Appendix \ref{app:eta_cc} \\ \hline
  $\eta_{tt}$ & $0.5765(65)$ 
  &\cite{Buras2008:PhysRevD.78.033005} \\ \hline
  $\eta_{ct}$ & $0.496(47)$
  &\cite{Brod2010:prd.82.094026}
  \end{tabular}
  \end{ruledtabular}
\end{table}

%
The remaining input parameters are the Fermi constant $G_F$, $W$ boson
mass $M_W$, quark masses $m_q$, kaon mass $m_{K^0}$, mass difference
$\Delta M_K$, and kaon decay constant $F_K$.
These are summarized in Table~\ref{tbl:in-others}.
\begin{table}[!th]
  \caption{Other Input Parameters}
  \label{tbl:in-others}
  \renewcommand{\arraystretch}{1.2}
  \begin{ruledtabular}
  \begin{tabular}{clc}
  Input & Value & Ref. \\ \hline\hline
  $G_{F}$ 
	& $1.1663787(6) \times 10^{-5}$ GeV$^{-2}$ 
  &\cite{Agashe2014:ChinPhysC.38.090001} \\ \hline
  $M_{W}$ 
	& $80.385(15)$ GeV 
  &\cite{Agashe2014:ChinPhysC.38.090001} \\ \hline
  $m_{c}(m_{c})$ 
	& $1.275(25)$ GeV 
  &\cite{Agashe2014:ChinPhysC.38.090001} \\ \hline
  $m_{t}(m_{t})$ 
	& $163.3(2.7)$ GeV 
  &\cite{Alekhin2012:plb.716.214} \\ \hline
  $\theta$ 
	& $43.52(5)^{\circ}$ 
  &\cite{Agashe2014:ChinPhysC.38.090001} \\ \hline
  $m_{K^{0}}$ 
	& $497.614(24)$ MeV 
  &\cite{Agashe2014:ChinPhysC.38.090001} \\ \hline
  $\Delta M_{K}$ 
	& $3.484(6) \times 10^{-12}$ MeV 
  &\cite{Agashe2014:ChinPhysC.38.090001} \\ \hline
  $F_K$
  & $156.2(7)$ MeV 
  &\cite{Agashe2014:ChinPhysC.38.090001}
  \end{tabular}
  \end{ruledtabular}
\end{table}

\subsection{Error Estimate}
\label{ssec:err}
We use the Monte Carlo method to obtain the expectation value of
$\epsK$,
\begin{equation}
  \int d^{d}\mathbf{x}\, \rho(\mathbf{x}) \epsK(\mathbf{x})
  = \frac{1}{N_s} \sum_{i=1}^{N_s} \epsK(x_i) 
    + \mathcal{O}\left(\frac{1}{\sqrt{N_s}}\right) \,,
\end{equation}
where $\mathbf{x}$ is a sample vector of the input parameters that we
describe in the previous section.
We generate $N_s=100,000$ random sample vectors $\mathbf{x}$ that
follow the multivariate Gaussian probability distribution
$\rho(\mathbf{x})$ with covariance matrix $C_{ij} = \expv{\delta x_i
  \delta x_j}, \delta x_i = x_i-\expv{x_i}$,
\begin{equation}
  \rho(\mathbf{x}) 
  = \mathcal{N} \exp{\Big( - \frac{1}{2}
    (\mathbf{x}-\expv{\mathbf{x}})^{T} C^{-1}
    (\mathbf{x}-\expv{\mathbf{x}}) \Big)} \,,
\end{equation}
where $\mathcal{N}$ is the probability density normalization factor.
The dimension of a sample vector $\mathbf{x}$ is $d=18$, which is the
total number of input parameters which appear in the master formula
for $\epsK$, Eq.~\eqref{eq:epsK_SM_0}.
We construct the covariance matrix by assuming a correlation $c_{ij}$
between parameters $x_i$ and $x_j$
\begin{equation}
  C_{ij} = c_{ij} \sigma_i \sigma_j \,,\; (-1 \leq c_{ij} \leq 1) \,,
\end{equation}
and using the mean $\expv{x_i}$ and error $\sigma_i$ of the input
parameter $x_i$ given in Tables~\ref{tbl:in-wolf}, \ref{tbl:in-Vcb},
\ref{tbl:in-BK}, \ref{tbl:in-qcdeta}, \ref{tbl:in-others},
Eq.~\eqref{eq:xi_0:rbc}, and Eq.~\eqref{eq:xi_LD:2}.
When the quoted error is asymmetric, we take the larger one as a
symmetric error.
The actual values of the correlation matrix $c_{ij}$ are given in 
Section \ref{sec:result}.

In this numerical study, we used the GNU Scientific Library
(GSL)~\cite{GSL:web}.
Specifically, we used the pseudo random number generator {\tt
  ranlxd2}~\cite{Luscher1993:Comput.Phys.Commun.79.100} to obtain
uniformly distributed random numbers.
Then we convert them to the multivariate Gaussian distribution using
GSL built-in functions.

To find the contribution to the total error from the error in each
parameter entering the master formula for $\epsK$, we use the
following error propagation method.
For $f = \abs{\epsK}$, the variance is
\begin{equation}
\sigma_{f}^{2} 
  = \expv{ [f(\mathbf{x})-f(\expv{\mathbf{x}})]^2 } 
  = \expv{[\delta f(\mathbf{x})]^{2}} \,,
\end{equation}
where
\begin{equation}
  \delta f(\mathbf{x}) 
  = \sum_{j=1}^{N} \frac{\partial f(\mathbf{x})}{\partial x_{j}} 
          \bigg\vert_{\expv{\mathbf{x}}} \delta x_{j} \,,
\end{equation}
as a linear approximation.
Then, the square of the relative error is obtained by 
\begin{align}
  \label{eq:errprop}
  \frac{\sigma_{f}^{2}}{\expv{f}^2} 
  &\approx \sum_{j,k=1}^{N} 
		   \frac{\partial f(\mathbf{x})}{\partial x_{j}} 
		   \bigg\vert_{\expv{\mathbf{x}}} 
		   \frac{\partial f(\mathbf{x})}{\partial x_{k}}
		   \bigg\vert_{\expv{\mathbf{x}}} 
       \frac{\expv{\delta x_{j} \delta x_{k}}}{\expv{f}^2} \CL
  &= \sum_{j,k=1}^{N} c_{jk}  
	 \cdot \frac{\partial f(\mathbf{x})}{\partial x_{j}}
	 \bigg\vert_{\expv{\mathbf{x}}} \frac{\sigma_{j}}{\expv{f}} 
	 \cdot \frac{\partial f(\mathbf{x})}{\partial x_{k}}
	 \bigg\vert_{\expv{\mathbf{x}}} \frac{\sigma_{k}}{\expv{f}} \,,
\end{align}
where $c_{ij}$ is again the correlation matrix; by definition the
diagonal components are always $c_{ii} = 1$.
This method of error propagation is used to cross-check our Monte
Carlo result.
Indeed, errors estimated by these two different methods are consistent
with each other.
And for the error budget in Table \ref{tbl:epsK-budget}, we quote the
fractional error for the parameter $x_i$, which is defined as
\begin{equation}
  \label{eq:errpropbudget}
  \bigg( \frac{\partial f(\mathbf{x})}{\partial x_{i}}
  \bigg\vert_{\expv{\mathbf{x}}} \frac{\sigma_{i}}{\expv{f}} 
  \bigg)^2 \bigg/
  \frac{\sigma_{f}^{2}}{\expv{f}^2} \,,
\end{equation}
in percent.

\section{Results} 
\label{sec:result}

Let us define $\epsK^\text{SM}$ as the theoretical evaluation of
$\abs{\eps_K}$ obtained using the master formula,
Eq.~\eqref{eq:epsK_SM_0}.
We define $\epsK^\text{Exp}$ as the experimental value of
$\abs{\eps_K}$, given in Eq.~\eqref{eq:epsK_exp}.
Let us define $\Delta\epsK$ as the difference between
$\epsK^\text{Exp}$ and $\epsK^\text{SM}$:
\begin{equation}
  \Delta \epsK \equiv \epsK^\text{Exp} - \epsK^\text{SM} \,.
  \label{eq:DepsK}
\end{equation}
Here, we assume that the theoretical phase $\theta$ in
Eq.~\eqref{eq:epsK_SM_0} is equal to the experimental phase
$\phi_\eps$ in Eq.~\eqref{eq:epsK_exp} \cite{
  Agashe2014:ChinPhysC.38.090001}.

\begin{figure*}[!t]
  \subfigure[CKMfitter]{%
    \includegraphics[width=0.31\textwidth]{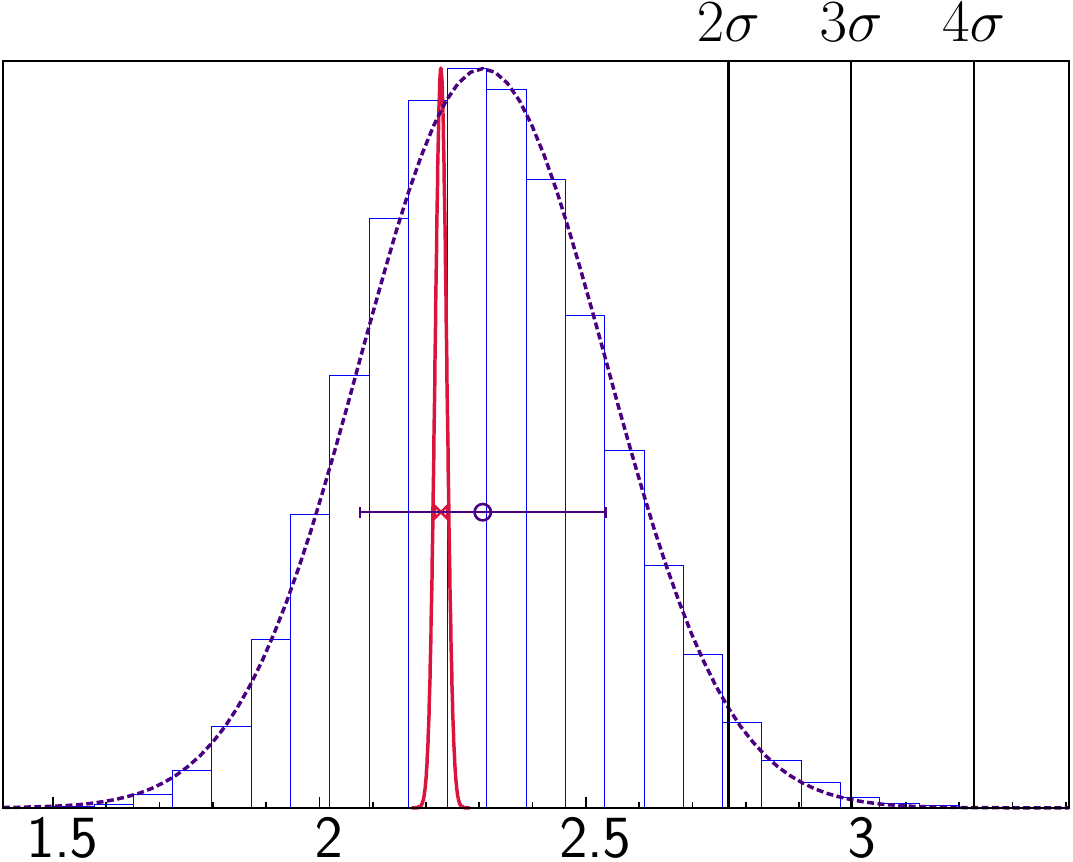}
    \label{sfig:incl.avg.CKM}
  }
  \hfill
  \subfigure[UTfit]{%
    \includegraphics[width=0.31\textwidth]{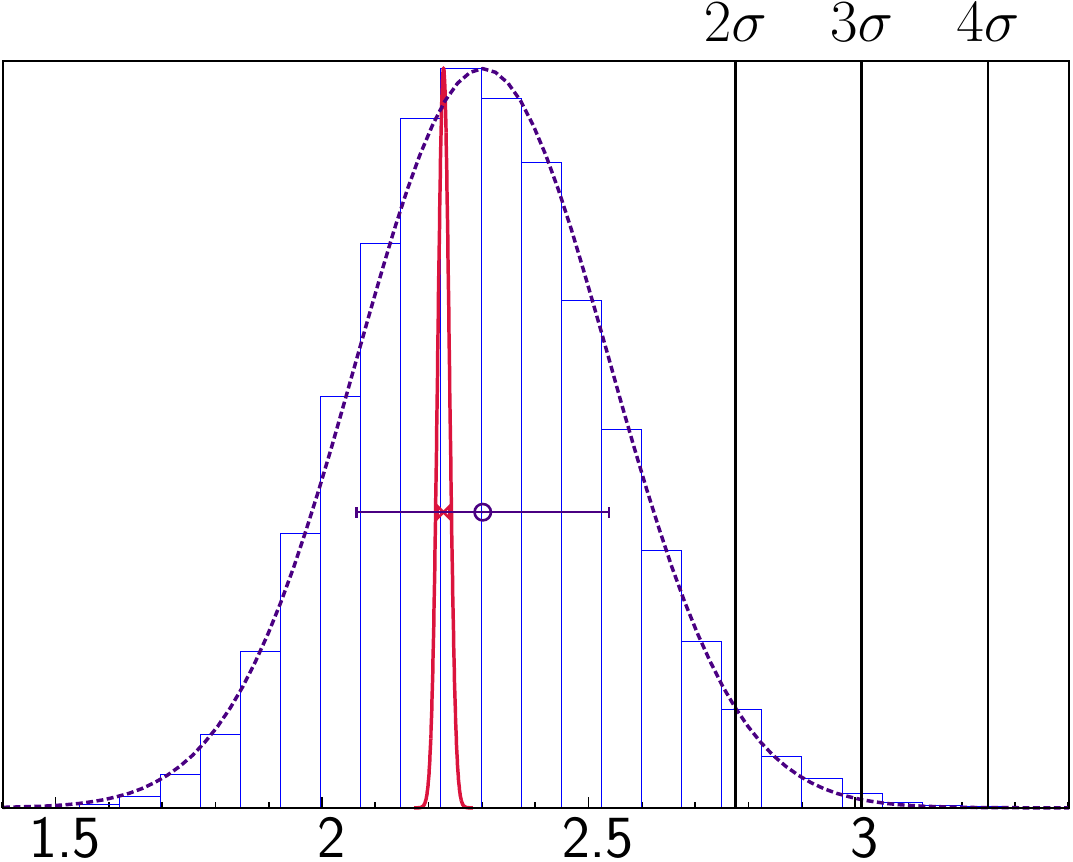}
    \label{sfig:incl.avg.UT}
  }
  \hfill
  \subfigure[AOF]{%
    \includegraphics[width=0.31\textwidth]{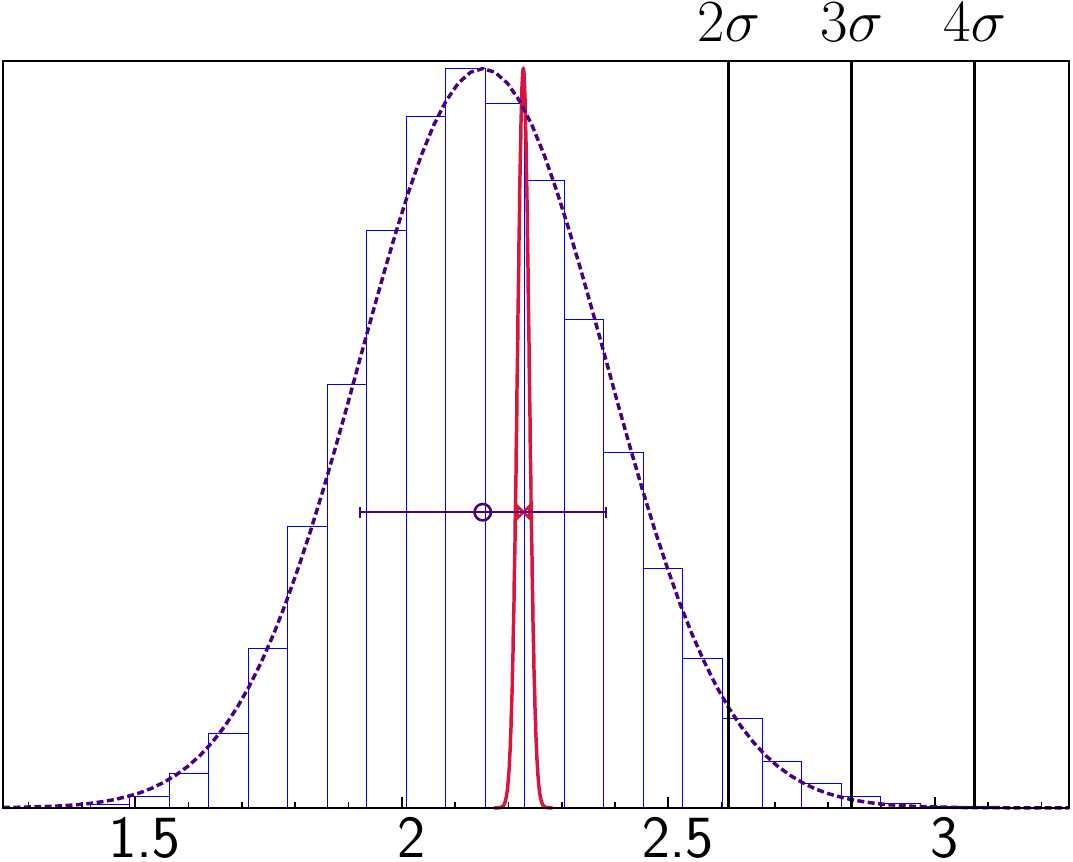}
    \label{sfig:incl.avg.AO}
  }
  \\
  \subfigure[CKMfitter]{%
    \includegraphics[width=0.31\textwidth]{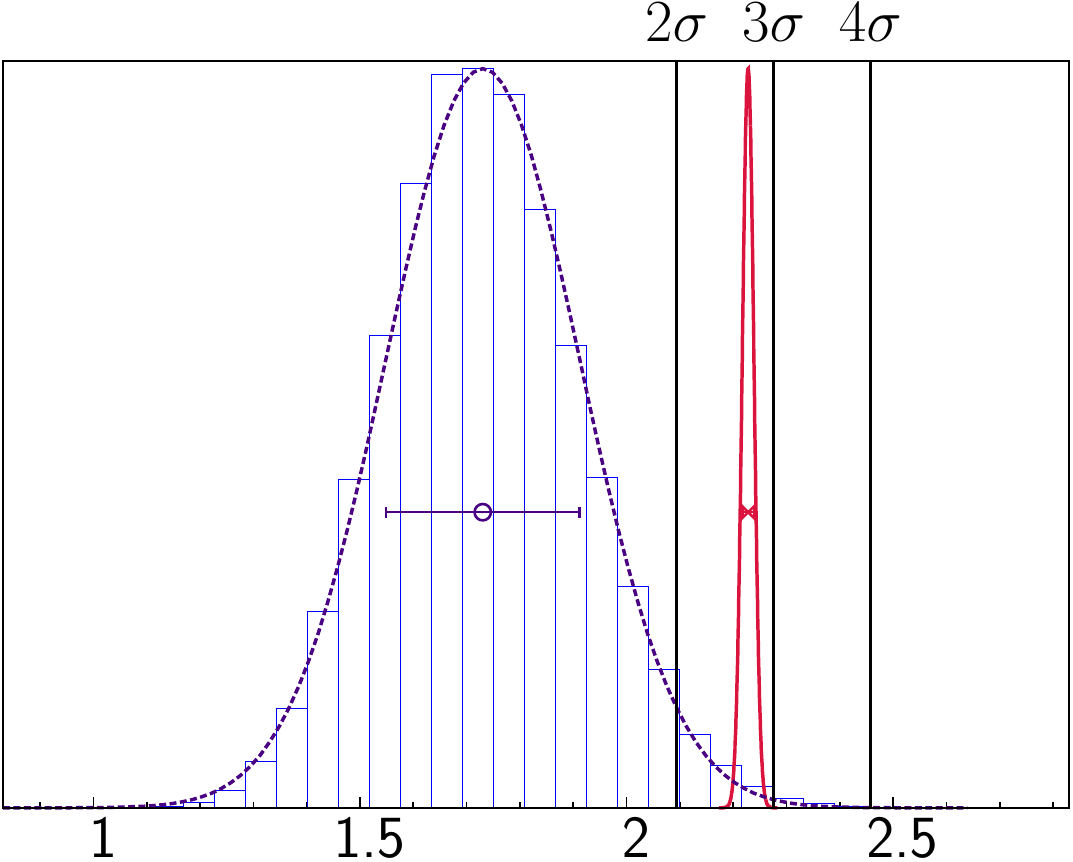}
    \label{sfig:excl.avg.CKM}
  }
  \hfill
  \subfigure[UTfit]{%
    \includegraphics[width=0.31\textwidth]{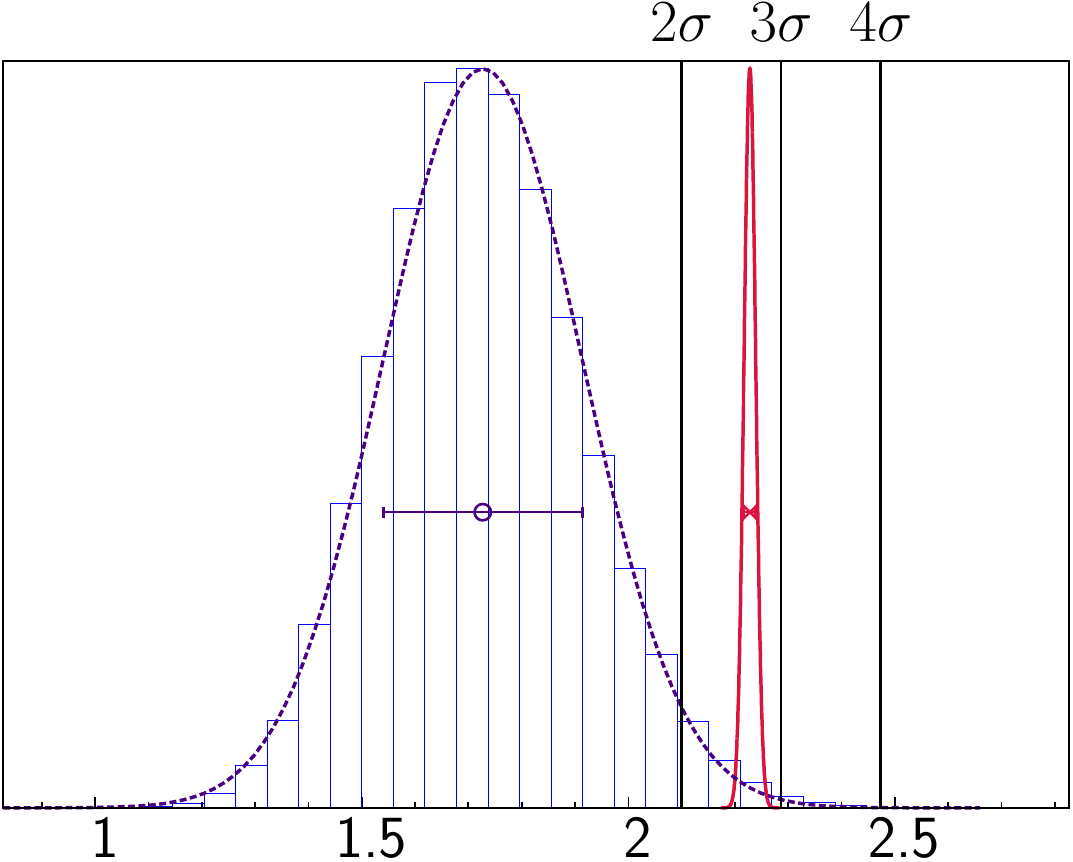}
    \label{sfig:excl.avg.UT}
  }
  \hfill
  \subfigure[AOF]{%
    \includegraphics[width=0.31\textwidth]{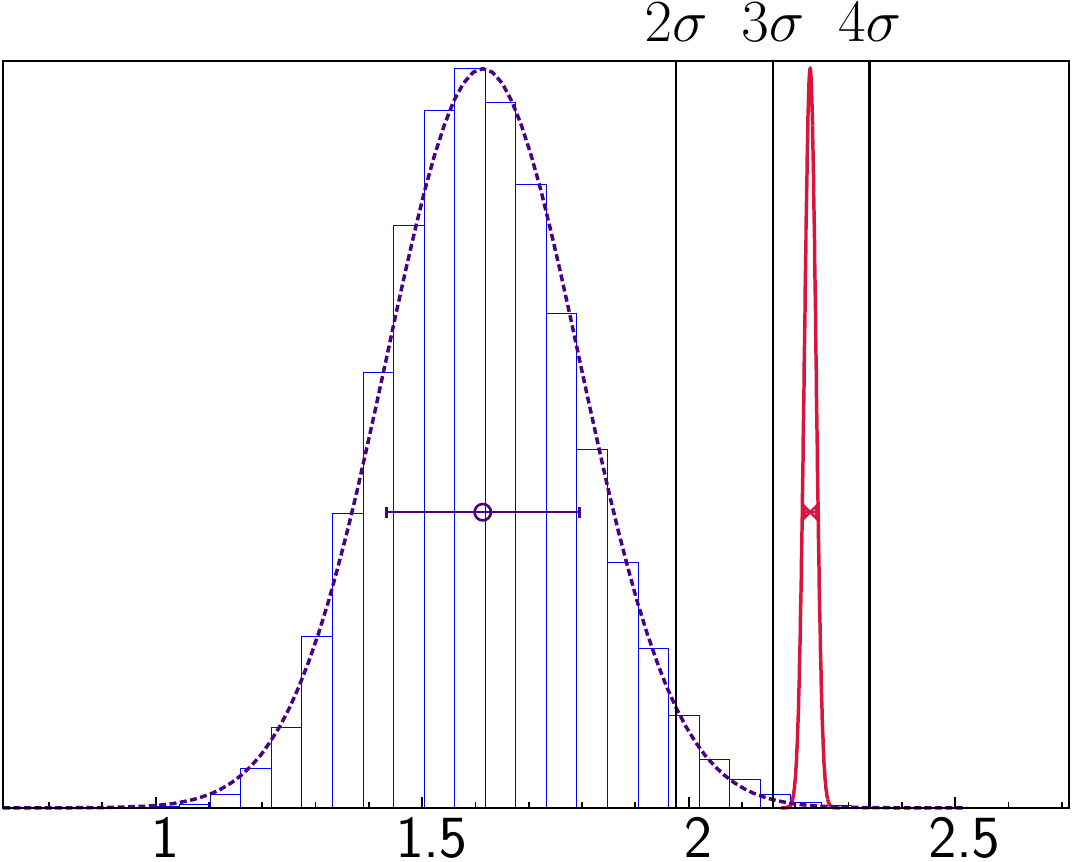}
    \label{sfig:excl.avg.AO}
  }
  \caption{Gaussian probability distributions for $\epsK^\text{SM}$
    (blue dotted line) and $\epsK^\text{Exp}$ (red solid line).  Here,
    the results are obtained using the FLAG $\BK$.  The results of
    \ref{sfig:incl.avg.CKM}, \ref{sfig:incl.avg.UT}, and
    \ref{sfig:incl.avg.AO} are obtained using the inclusive $\Vcb$.
    Those of \ref{sfig:excl.avg.CKM}, \ref{sfig:excl.avg.UT}, and
    \ref{sfig:excl.avg.AO} are obtained using the exclusive $\Vcb$. }
  \label{fig:hstgwFLAG}
\end{figure*}

In Table \ref{tbl:epsKwFLAG}, we present results for $\epsK^\text{SM}$
obtained using the FLAG average for $\BK$ \cite{
  Aoki2013:hep-lat.1310.8555} and $V_{cb}$ from both inclusive \cite{
  Alberti2014:PhysRevLett.114.061802} and exclusive channels \cite{
  Bailey2014:PhysRevD.89.114504}.
The corresponding probability distributions for $\epsK^\text{SM}$ and
$\epsK^\text{Exp}$ are presented in Fig.~\ref{fig:hstgwFLAG}.
The corresponding results for $\Delta \epsK$ are presented in
Table \ref{tbl:DepsKwFLAG}.
%
\begin{table}[!tbh]
  \caption{$\epsK^\text{SM}$ in units of $10^{-3}$. Here, we use the
    FLAG average for $\BK$ in Table \ref{tbl:in-BK}.  The input
    methods of CKMfitter, UTfit, and AOF represent different inputs
    for the Wolfenstein parameters, which are explained in detail in
    Section \ref{ssec:iparam}. 
  }
  \label{tbl:epsKwFLAG}
  \renewcommand{\arraystretch}{1.2}
  \begin{ruledtabular}
  \begin{tabular}{ccc}
  Input Method & Inclusive $\Vcb$ & Exclusive $\Vcb$ 
  \\ \hline
  CKMfitter 
  & $2.31(23)$ 
  & $1.73(18)$ 
  \\ \hline
  UTfit
  & $2.30(24)$ 
  & $1.73(19)$ 
  \\ \hline
  AOF
  & $2.15(23)$ 
  & $1.61(18)$ 
  \end{tabular}
  \end{ruledtabular}
\end{table}
\begin{table}[!tbh]
  \caption{$\Delta \epsK$. Here, we use $\epsK^\text{SM}$ from
    Table~\ref{tbl:epsKwFLAG}. We obtain $\sigma$ by combining
    the errors of $\epsK^\text{SM}$ and $\epsK^\text{Exp}$ in
    quadrature.}
  \label{tbl:DepsKwFLAG}
  \renewcommand{\arraystretch}{1.2}
  \begin{ruledtabular}
    \begin{tabular}{ccc}
  Input Method & 
  Inclusive $\Vcb$ & 
  Exclusive $\Vcb$
  \\ \hline
  CKMfitter 
  & $-0.34\sigma$ 
  & $2.7\sigma$ 
  \\ \hline
  UTfit
  & $-0.31\sigma$ 
  & $2.7\sigma$ 
  \\ \hline
  AOF
  & $0.33\sigma$ 
  & $3.4\sigma$ 
  \end{tabular}
  \end{ruledtabular}
\end{table}

From Table \ref{tbl:DepsKwFLAG}, we find that $\epsK^\text{SM}$ with
inclusive $V_{cb}$ is consistent with $\epsK^\text{Exp}$ within
$1\sigma$.
In other words, $\Delta \epsK$ is consistent with zero with inclusive
$V_{cb}$ regardless of the input methods.

However, from Tables \ref{tbl:epsKwFLAG} and \ref{tbl:DepsKwFLAG},
$\epsK^\text{SM}$ with exclusive $V_{cb}$ is only 72\% of
$\epsK^\text{Exp}$.
For this case, with the most reliable input method (AOF),
$\Delta\epsK$ is $3.4\sigma$.
Since the largest contribution in our estimate of $\epsK^\text{SM}$
that we neglect is much less than $2\%$, the neglected contributions
cannot explain the gap $\Delta\epsK$ of 28\% with exclusive $V_{cb}$.
Hence, our final results for $\Delta\epsK$ are
\begin{align}
\Delta\epsK &= 3.4 \sigma &\quad &\text{(exclusive $V_{cb}$)}\,,
\label{eq:final-DepsK:ex}
\\
\Delta\epsK &= 0.33 \sigma &\quad &\text{(inclusive $V_{cb}$)}\,,
\label{eq:final-DepsK:in}
\end{align}
where we take the AOF result as our quoted value.

In the case of the FLAG $\BK$, the BMW result for $\BK$ \cite{
  Durr2011:PhysLettB.705.477} dominates the FLAG result, and the gauge
ensembles used for the BMW calculation are independent of those used
for the determination of exclusive $V_{cb}$ \cite{
  Bailey2014:PhysRevD.89.114504} by the FNAL/MILC collaboration.
Hence, we assume that we may neglect the correlation between the FLAG
$\BK$ and the exclusive $V_{cb}$.
However, the SWME $\BK$ calculation in Ref.~\cite{
  Bae2014:prd.89.074504} shares the same MILC gauge ensembles with the
exclusive $V_{cb}$ determination in Ref.~\cite{
  Bailey2014:PhysRevD.89.114504}.
Hence, in this case, we cannot neglect the correlation between
the SWME $\BK$ and the exclusive $V_{cb}$.
We introduce $+50\%$ correlation and $-50\%$ anti-correlation between
the SWME $\BK$ and the exclusive $V_{cb}$, and take the
maximum deviation from the uncorrelated case as the systematic error
due to the unknown correlation between them.
The details of this analysis are explained in Appendix
\ref{app:epsKwSWME}.
However, this analysis shows that the size of the ambiguity due to the
correlation between the SWME $\BK$ and the exclusive $V_{cb}$ is so
large that we can use the results of the SWME $\BK$ only to
cross-check those with the FLAG $\BK$.
This analysis of the correlation is another update from the previous
paper \cite{Bailey:2014qda}.

\begin{figure}[!tbh]
  \centering
  \includegraphics[width=0.9\columnwidth]{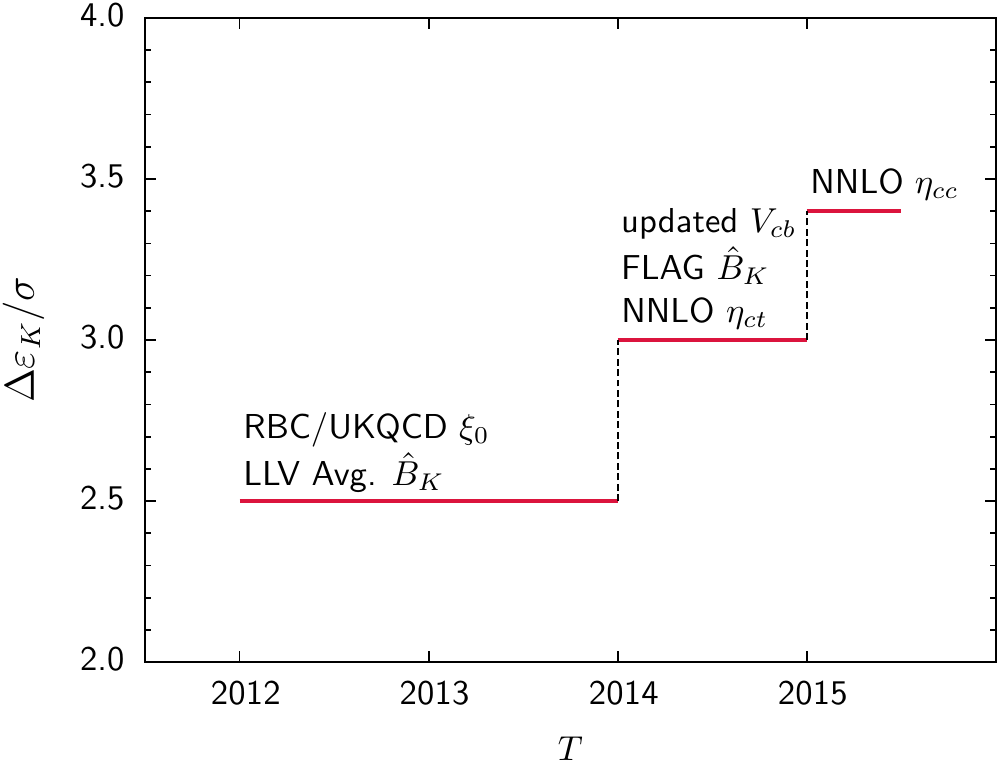}
  \caption{Recent history of $\Delta \epsK$ along with the theoretical
    progress.}
  \label{fig:history}
\end{figure}
It is interesting to understand the historical evolution of $\Delta
\epsK/\sigma$ along with the theoretical progress in lattice QCD
and perturbative QCD.
In Fig.~\ref{fig:history}, we present $\Delta \epsK/\sigma$ as a
function of time.
In 2012, the RBC/UKQCD collaboration reported $\xi_0$ in Ref.~\cite{
  Blum2011:PhysRevLett.108.141601}.
In addition to this, using the LLV average for $\BK$ \cite{
  Laiho2009:PhysRevD.81.034503}, the SWME collaboration reported
$\Delta\epsK = 2.5\sigma$ in Ref.~\cite{ Jang2012:PoS.LAT2012.269}
in 2012.
In 2014, FNAL/MILC reported an updated $V_{cb}$ in the exclusive
channel.
Using the FLAG average for $\BK$ \cite{ Aoki2013:hep-lat.1310.8555}
and the NNLO value of $\eta_{ct}$ \cite{Brod2010:prd.82.094026}, the
SWME collaboration reported the updated $\Delta\epsK = 3.0\sigma$ in
Ref.~\cite{ Bailey:2014qda} in 2014.\footnote{Here, we evaluate the
  $\epsK$ using the correct master formula in Eq.~\eqref{eq:epsK_SM_0}
  with the same inputs as in \cite{ Jang2012:PoS.LAT2012.269} (2012)
  and \cite{ Bailey:2014qda} (2014).}
In this paper, we investigate issues in the NNLO calculation of
$\eta_{cc}$ \cite{ Brod2011:PhysRevLett.108.121801,
  Buras2013:EurPhysJC.73.2560} and use the SWME result in
Table~\ref{tbl:in-qcdeta} to report the updated $\Delta\epsK =
3.4\sigma$ in Eq.~\eqref{eq:final-DepsK:ex}.
%

\section{Conclusion}
\label{sec:conclude}

In this paper, we observe that there is a substantial $3.4\sigma$
tension in $\epsK$ between experiment and the SM theory with lattice
QCD inputs.
For this claim, we choose the angle-only fit (AOF), the exclusive
$V_{cb}$ (lattice QCD results), and the FLAG $\BK$ (lattice QCD
results) to determine the final value.
We choose the AOF method to determine the final result because the
AOF Wolfenstein parameters do not have unwanted correlation with
$\epsK$ and $\BK$.
However, the tension disappears in the case of inclusive $V_{cb}$
(results of the heavy quark expansion based on the OPE) regardless of
the choices for the Wolfenstein parameters.

In Table \ref{tbl:epsK-budget}, we present the error budget of
$\epsK^\text{SM}$ for the central value.
This is obtained using the error propagation method explained in
Section \ref{ssec:err}.
From this error budget, we find out that $V_{cb}$ dominates the error
in $\epsK^\text{SM}$.
Hence, it is essential to reduce the error of $V_{cb}$ as much as
possible. (See also Refs.~\cite{ Buras:2014sba, Buras:2013ooa}.)
In order to achieve this goal, we plan to extract $V_{cb}$ from the
exclusive channel using the Oktay-Kronfeld (OK) action \cite{
  Oktay2008:PhysRevD.78.014504} for heavy quarks to calculate the form
factors for $\bar{B} \to D^{(*)} \ell \bar{\nu}$ decays.
Preliminary results in the early stage of the $V_{cb}$ project are
reported in Refs.~\cite{ Jang:LAT2013, Bailey:LAT2014, Jang:LAT2014}.
%
\begin{table}[tb!]
  \caption{ Error budget for $\epsK^\text{SM}$ obtained using the AOF
    method, the exclusive $\Vcb$, and the FLAG $\BK$. Here, the values
    are fractional contributions to the total error obtained using the
    formula in Eq.~\eqref{eq:errpropbudget}.}
  \label{tbl:epsK-budget}
  \begin{ruledtabular}
  \begin{tabular}{ccc}
    source       & error (\%) & memo \\
    \hline
    $V_{cb}$     & 39.3        & FNAL/MILC \\
    $\bar{\eta}$ & 20.4        & AOF \\
    $\eta_{ct}$  & 16.9        & $c-t$ Box \\
    $\eta_{cc}$  &  7.1        & $c-c$ Box \\
    $\bar{\rho}$ &  5.4        & AOF \\
    $m_t$        &  2.4        & \\
    $\xi_0$      &  2.2        & RBC/UKQCD\\
    $\xi_\text{LD}$      &  2.0        & RBC/UKQCD\\
    $\hat{B}_K$  &  1.5        & FLAG \\
    $m_c$        &  1.0        & \\
    $\vdots$     & $\vdots$    & 
  \end{tabular}
  \end{ruledtabular}
\end{table}
%
%
%

Our results for $\eta_{cc}$ are consistent with the conclusion of
Ref.~\cite{ Brod2011:PhysRevLett.108.121801} regarding the convergence
of perturbation theory.
Uncertainty due to truncated higher order terms requires further
investigation in the future.
Lattice QCD calculations with dynamical charm quarks, such as that
envisioned by the RBC/UKQCD collaboration, could shed light on this
issue.

We expect that our results for $\epsK$ would be consistent with those
from a global UT analysis, such as that in Ref.~\cite{
  Laiho2009:PhysRevD.81.034503}.
The authors of Ref.~\cite{ Laiho2009:PhysRevD.81.034503} performed one
analysis with inclusive $V_{cb}$ and another analysis with exclusive
$V_{cb}$ instead of inflating their errors and taking the average.
In this respect, the analysis of Ref.~\cite{
  Laiho2009:PhysRevD.81.034503} is different from those of UTfit and
CKMfitter.
Such a global analysis with up-to-date inputs from lattice QCD has not
been performed yet.
It would be interesting to see the results of such an analysis.

\begin{acknowledgments}
Y.C.J. thanks to Amarjit Soni for helpful discussion on the unitarity
triangle analysis.
We thank to J.~Brod and A.~J. Buras for a useful discussion on the
$\eta_{cc}$.
The research of W.~Lee is supported by the Creative Research
Initiatives Program (No.~2014001852) of the NRF grant funded by the
Korean government (MEST).
W.~Lee would like to acknowledge the support from the KISTI
supercomputing center through the strategic support program for the
supercomputing application research [No.~KSC-2014-G3-002].
Computations were carried out on the DAVID GPU clusters at Seoul
National University.
J.A.B. is supported by the Basic Science Research Program of the
National Research Foundation of Korea (NRF) funded by the Ministry of
Education (No.~2014027937).
\end{acknowledgments}

\appendix
\section{Next-to-next-to leading order $\eta_{cc}$}
\label{app:eta_cc}

We will begin from the master formula for $\eta_{cc}$ \cite{
  Brod2011:PhysRevLett.108.121801} and give an explicit expression for
each component which is necessary for a numerical evaluation.
For $\mu \leq \mu_c$, 
\begin{equation}
  \eta_{cc} 
  = \frac{1}{m_c^2 \left(m_c\right)} \tilde C_{S2}^{cc}\left(\mu_c\right)
  \left[\alpha_s\left(\mu_c,3\right)\right]^{a_{+}(3)}
  K_{+}^{-1}(\mu_c,3) \,.
\end{equation}

The magic number $a_+(3) = 2/9$ can be obtained from Eq.~\eqref{eq:adm0}.
$\alpha_s(\mu,f)$ is the running strong coupling constant with $f$
active flavors at scale $\mu$.
We will use the four-loop $\alpha_s$ running formula \cite{
  Chetyrkin2000:CompPhysComm.133.43,
  Chetyrkin1997:PhysRevLett.79.2184}.
The Wilson coefficient $\tilde{C}_{S2}^{cc}(\mu_c)$ of the $\Delta
S=2$ four-fermion operator is defined by Eq.~\eqref{eq:match_at_mu_c}.
%
%
The running matrix $K_{+}^{-1}(\mu_c,3)$ is given by
Eq.~\eqref{eq:rg_K_inv-2}.

At the charm scale $\mu_c$, the effective four flavor theory is
matched to the effective three flavor theory by requiring the
following condition \cite{Brod2011:PhysRevLett.108.121801},
\begin{equation}
  \label{eq:match_at_mu_c}
  \sum_{i,j=+,-} C_i C_j \langle Q_{i} Q_j
  \rangle = \frac{1}{8\pi^2} {\tilde C}_{S2}^{cc}
  \langle {\tilde Q}_{S2} \rangle \,.
\end{equation}
The matrix elements and the Wilson coefficients are expanded in the
three flavor strong coupling $\alpha_s(\mu_c,3)$,
\begin{align}
  \expv{{\tilde Q}_{S2}}
  =& \tilde{r}_{S2} \expv{{\tilde Q}_{S2}}^{(0)} \,,\\
  \tilde{r}_{S2} =& 1 + \frac{\alpha_s(\mu_c,3)}{4\pi} \tilde{r}_{S2}^{(1)}
            + \left(\frac{\alpha_s(\mu_c,3)}{4\pi}\right)^2 \tilde{r}_{S2}^{(2)} \,,\\
  \expv{Q_i Q_j} 
  =& \frac{m_c^2(\mu_c)}{8\pi^2} d_{ij} \expv{{\tilde Q}_{S2}}^{(0)} \,,\\
  d_{ij}  
  =&  d_{ij}^{(0)} + \frac{\alpha_s(\mu_c,3)}{4\pi} d_{ij}^{(1)}
     + \left(\frac{\alpha_s(\mu_c,3)}{4\pi}\right)^2 d_{ij}^{(2)} \,,
\end{align}
\begin{align}
  \label{eq:expandCif3}
  C_{i}(\mu_c) 
  =& C_i^{(0)}(\mu_c)
   + \frac{\alpha_s(\mu_c,3)}{4\pi} C_i^{(1)}(\mu_c) \CL
   &+ \left(\frac{\alpha_s(\mu_c,3)}{4\pi}\right)^2 C_i^{(2)}(\mu_c) \,,\\
 \tilde{C}_{S2}^{cc}(\mu_c)
  =& \tilde{C}_{S2}^{cc(0)}(\mu_c)
   + \frac{\alpha_s(\mu_c,3)}{4\pi} \tilde{C}_{S2}^{cc(1)}(\mu_c) \CL
   &+ \left(\frac{\alpha_s(\mu_c,3)}{4\pi}\right)^2 \tilde{C}_{S2}^{cc(2)}(\mu_c) \,.
\end{align}
Then, the matching results are
\begin{align}
  \label{eq:match_result-1}
  \tilde{C}_{S2}^{cc(0)}(\mu_c) 
  =& m_c^2(\mu_c)
  C_{i}^{(0)} C_{j}^{(0)} d_{ij}^{(0)} \, , \\
  \tilde{C}_{S2}^{cc(1)}(\mu_c) 
  =& m_c^2(\mu_c) \bigg[
  C_i^{(0)} C_j^{(0)} \hat{d}_{ij}^{(1)} \CL
  &+ \big( C_i^{(1)} C_j^{(0)} +
  C_i^{(0)} C_j^{(1)} \big) d_{ij}^{(0)} \bigg] \, , 
  \label{eq:match_result-2} \\ 
  \tilde{C}_{S2}^{cc(2)}(\mu_c) 
  =& m_c^2(\mu_c) \bigg[
  C_i^{(0)} C_j^{(0)} 
  \left(\hat{d}_{ij}^{(2)} + \frac{2}{3} \log \frac{\mu_c^2}{m_c^2} d_{ij}^{(1)} \right) \CL
  & \hspace*{-2pc} 
  + \big( C_i^{(1)} C_j^{(0)} + C_i^{(0)} C_j^{(1)} \big)
  \left( \hat{d}_{ij}^{(1)} + \frac{2}{3} \log \frac{\mu_c^2}{m_c^2} d_{ij}^{(0)} \right) \nonumber \\ 
  &+ \big( C_i^{(2)} C_j^{(0)} +
  C_i^{(1)} C_j^{(1)} +  C_i^{(0)} C_j^{(2)} \big) d_{ij}^{(0)}
  \bigg] \, ,
\label{eq:match_result-3}
\end{align}
where $m_c=m_c(m_c)$ in the logarithms multiplied by $d_{ij}^{(0,1)}$,
and
\begin{align}
  \hat{d}_{ij}^{(1)} 
  \equiv& d_{ij}^{(1)} - d_{ij}^{(0)}  \tilde r_{S2}^{(1)} \,,\\
  \hat{d}_{ij}^{(2)} 
  \equiv& d_{ij}^{(2)} - \hat{d}_{ij}^{(1)} \tilde{r}_{S2}^{(1)} 
         - d_{ij}^{(0)}\tilde r_{S2}^{(2)} \,.
\end{align}

Note that the matching scale is the charm quark mass $\mu_c=m_c(m_c)$;
in Eqs.~\eqref{eq:match_result-1}, \eqref{eq:match_result-2}, and
\eqref{eq:match_result-3}, the Wilson coefficients
$C_i^{(l)}(\mu_c)\ (l=0,1,2\,;i=\pm)$ are evaluated at $\mu_c=m_c$,
\begin{equation}
  \label{eq:Cpm_mc}
  C_{i}^{(l)} = C_{i}^{(l)}(m_c) \,.
\end{equation}
These are obtained by renormalization group evolution from the scale
$\mu_W$ down to the scale $\mu_c=m_c$.
(See Eq.~\eqref{eq:running_result}.)
%
To examine the size of residual scale dependence, we vary $\mu_c$,
keeping the condition Eq.~\eqref{eq:Cpm_mc}.

\begin{figure}[t!]
  \includegraphics[width=.85\columnwidth]{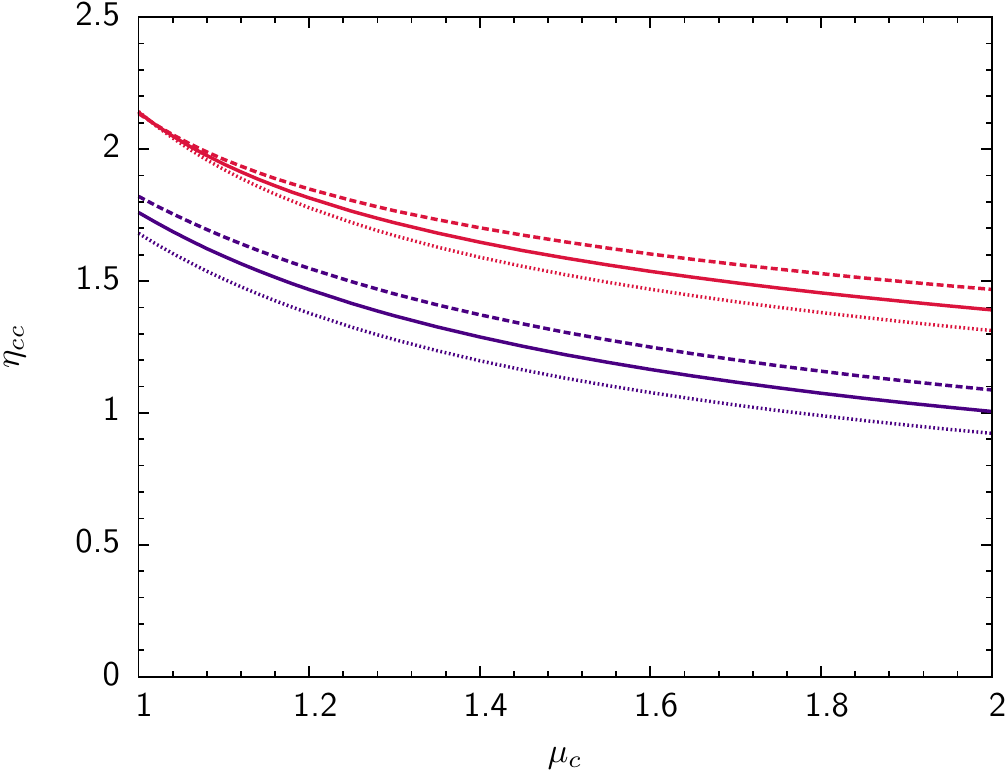}
  \caption{Scale $\mu_c$ dependence of $\eta_{cc}$. Dotted, solid, and
    dashed lines represent results for $\mu_W=40,\ 80$, and
    $160\;\GeV$, respectively.  The upper three (red) lines are the
    NNLO results, and the lower three (purple) lines are the NLO
    results.}
  \label{fig:etacc}
\end{figure}

Then the residual scale $\mu_c$ dependence in
$\tilde{C}_{S2}^{(l)}(\mu_c)$ enters from logarithms which are shown
explicitly in Eq.~\eqref{eq:match_result-3} and through $d_{ij}^{(l)}$
and $\tilde{r}_{S2}^{(l)}$; it also comes from the expansion
$m_c(\mu_c)$.
The expansion of the charm quark mass $m_c(\mu_c)$ near $\mu_c=m_c$ is
given by Eq.~\eqref{eq:running_m_q} with $f=4$.
The resulting residual scale dependence in $\eta_{cc}$ can be seen
from Fig.~\ref{fig:etacc}.

The leading and next-to-leading order (NLO) calculations can be found
from Ref.~\cite{Herrlich1993:NuclPhysB.419.292}, with the number of
colors $N_c=3$, $l_c =
\log\left(\mu_c^2/m_c^2\left(\mu_c\right)\right)$,
\begin{gather}
  \label{eq:coef_d}
  d_{++}^{(0)} = \frac{3}{2} \,,\CL
  d_{+-}^{(0)} = d_{-+}^{(0)} = -\frac{1}{2} \,,\CL
  d_{--}^{(0)} = \frac{1}{2} \,,
\end{gather}
\begin{gather}
  d_{++}^{(1)} = 9 l_c - \frac{27}{2} - \frac{\pi^2}{6} \,,\CL
  d_{+-}^{(1)} = d_{-+}^{(1)} 
  = -6 l_c -\frac{23}{6} + \frac{5\pi^2}{18} \,,\CL
  d_{--}^{(1)} = 6 l_c + \frac{53}{6} + \frac{\pi^2}{18} \,,
\end{gather}
\begin{equation}
  \tilde{r}_{S2}^{(1)} = -\frac{17}{3} \,.
\end{equation}

The next-to-next-to-leading order (NNLO) calculation results are
presented in Ref.~\cite{Brod2011:PhysRevLett.108.121801},
\begin{align}
 \hat d_{++}^{(2)} 
 =& \frac{1665873233}{8164800}-\frac{1573}{162} B_4-\frac{133}{72} D_3 \CL
  &+\frac{49}{36}\zeta_2 l_c +\frac{4313}{216} l_c^2 -\frac{15059}{1296} l_c \CL
  &+\frac{210213}{560} S_2 -\frac{1501}{54} \zeta_2^2
   -\frac{7567241}{204120} \zeta_2 \CL 
  &-\frac{1697893}{7776} \zeta_3 +\frac{11575}{216} \zeta_4 \,,
\end{align}
\begin{align}
  \hat d_{+-}^{(2)} = \hat d_{-+}^{(2)} 
  =& \frac{87537463}{1166400}+\frac{685}{162} B_4-\frac{83}{72} D_3 \CL
   &+\frac{695}{36}\zeta_2 l_c -\frac{1475}{216} l_c^2
    -\frac{57763}{1296}l_c \CL
   &-\frac{4797}{80} S_2
    -\frac{791}{54} \zeta_2^2+\frac{366569}{29160} \zeta_2 \CL
   &+\frac{57673}{7776} \zeta_3 -\frac{4999}{216} \zeta_4 \,,
\end{align}
\begin{align}
  \hat d_{--}^{(2)}
  =& \frac{2129775941}{8164800} + \frac{491}{162} B_4+\frac{11}{72} D_3 \CL
   &+\frac{865}{36} \zeta_2 l_c +\frac{12533}{216} l_c^2
    +\frac{171121}{1296}l_c \CL
   &+\frac{59121}{560} S_2  -\frac{517}{54} \zeta_2^2
    +\frac{9261883}{204120} \zeta_2 \CL
   &-\frac{411709}{7776} \zeta_3 -\frac{7913}{216} \zeta_4 \,.
\end{align}

Some constants for the master integrals are \cite{
  Steinhauser2000:CompPhysComm.134.335}
\begin{align}
  D_3
  =& 6\zeta_3 - \frac{15}{4}\zeta_4 
    - 6 \left[\mathrm{Cl}_2\left(\frac{\pi}{3}\right)\right]^2 \,,\CL
  B_4
  =& -4\zeta_2 \ln^2{2} + \frac{2}{3} \ln^4{2} - \frac{13}{2}\zeta_4
    + 16 \mathrm{Li}_4\left(\frac{1}{2}\right) \,,\CL
  S_2
  =& \frac{4}{9\sqrt{3}} \mathrm{Cl}_2\left(\frac{\pi}{3}\right) \,,
\end{align}
with
\begin{align}
  \label{eq:cl2}
  \mathrm{Cl}_2(x) 
  =& \mathrm{Im} \left( \mathrm{Li}_2 ( e^{ix} ) \right) \,,\\
  \mathrm{Li}_n(z)
  =& \sum_{k=1}^{\infty} \frac{z^k}{k^n} \,,
\end{align}
and the Riemann zeta function is
\begin{equation}
  \label{eq:zetan}
  \zeta_n = \sum_{k=1}^{\infty} \frac{1}{k^n} \,.
\end{equation}

In numerical evaluation, we use approximated numbers
which are obtained using \texttt{Mathematica}.
\begin{align}
  \label{eq:numeric_consts}
  \zeta_2 =& 1.644934 \ldots
          = \frac{\pi^2}{6} \,,\CL
  \zeta_3 =& 1.202056 \ldots \,,\CL
  \zeta_4 =& 1.082323 \ldots
          = \frac{\pi^4}{90} \,,\CL
  \mathrm{Li}_4\left(\frac{1}{2}\right) 
  =& 0.5174790 \ldots \,,\CL
  \mathrm{Cl}_2\left(\frac{\pi}{3}\right)
  =& 1.014941 \ldots \,.
\end{align}
For $\zeta_2$ and $\zeta_4$, we also give the exact expression.

The renormalization group evolution of the Wilson coefficients $C_\pm$
is described by the evolution matrix $U_{ij}$
\cite{Gorbahn2004:NuclPhysB.713.291},
\begin{equation}
  \label{eq:wilson_c_mu}
  C_{i}(\mu) = U_{ij}(\mu,\mu_0) C_{j}(\mu_0) \,,
\end{equation}
which is diagonalized by the specific choice of evanescent operators,
\begin{equation}
  \label{eq:evol_mat_U}
  U_{ij}(\mu,\mu_0) 
  = K_i(\mu) \left( \frac{\alpha_s(\mu_0,f)}{\alpha_s(\mu,f)} \right)^{a_i}
    K_i^{-1}(\mu_0) \delta_{ij} \,,
\end{equation}
where
\begin{align}
  \label{eq:rg_K_inv}
  K_\pm(\mu)
  =& 1 + \frac{\alpha_s(\mu,f)}{4\pi} J_\pm^{(1)} 
     + \Big( \frac{\alpha_s(\mu,f)}{4\pi} \Big)^2 J_\pm^{(2)} \,,\\
  \label{eq:rg_K_inv-2}
  K_\pm^{-1} (\mu_0)
  =& 1 - \frac{\alpha_s(\mu_0,f)}{4\pi} J_\pm^{(1)} \CL
   &- \Big( \frac{\alpha_s(\mu_0,f)}{4\pi} \Big)^2 \big( J_\pm^{(2)} - (J_\pm^{(1)})^2 \big) \,,
\end{align}
and
\begin{align}
  J_\pm^{(1)} 
  =& \frac{\beta_1}{\beta_0} a_\pm - \frac{\gamma_\pm^{(1)}}{2\beta_0} \,,\\
  J_\pm^{(2)}
  =& \frac{\beta_2}{2\beta_0} a_\pm
    + \frac{1}{2} \left( (J_\pm^{(1)})^2 
                       - \frac{\beta_1}{\beta_0} J_\pm^{(1)} \right)
    - \frac{\gamma_\pm^{(2)}}{4\beta_0} \,.
\end{align}
The expansion coefficients of the QCD beta function $\beta_i$ are given in Eq.~\eqref{eq:qcdbeta}.
The anomalous dimensions $\gamma_\pm^{(i)}$ for the operators $Q_\pm$ are taken from Ref.~\cite{Buras2006:JHEP.11.002},
\begin{align}
  \label{eq:adm0}
  \gamma_\pm^{(0)} & = \pm 6 \left ( 1 \mp \frac{1}{3} \right )
  = 2\beta_0 a_\pm \,,\\
  \gamma_\pm^{(1)} & = \left ( -\frac{21}{2} \pm \frac{2}{3} f \right )
  \left ( 1 \mp \frac{1}{3} \right ) \,,\\
  \gamma_\pm^{(2)} & = \frac{1}{300} \left ( 349049 \pm 201485 \right ) -
  \frac{1}{1350} \left ( 115577 \mp 9795 \right ) f \CL
  & \mp \frac{130}{27} \left ( 1 \mp \frac{1}{3} \right ) f^2 \mp \left (
  672 + 80 \left ( 1 \mp \frac{1}{3} \right ) f \right ) \zeta_3 \,.
\end{align}

The number of active flavors is fixed while applying
Eq.~\eqref{eq:wilson_c_mu}.
The number of flavors is implied by the strong coupling constant in
Eq.~\eqref{eq:evol_mat_U}.


The initial conditions for the Wilson coefficients $C_\pm$ are chosen
at the scale $\mu_W$,
\begin{align}
  C_\pm (\mu_W) 
  =& C_\pm^{(0)}(\mu_W) 
    + \frac{\alpha_s(\mu_W,5)}{4\pi} C_\pm^{(1)}(\mu_W) \CL
   &+ \left(\frac{\alpha_s(\mu_W,5)}{4\pi}\right)^2 C_\pm^{(2)}(\mu_W)
    \,.
\end{align}
The expansion coefficients are given in Ref.~\cite{
  Buras2006:JHEP.11.002},
\begin{align}
  \label{eq:ic}
  C_\pm^{(0)}(\mu_W) =& 1 \,,\CL
  C_\pm^{(1)}(\mu_W) 
  =& \pm \frac{1}{2} \left( 1 \mp \frac{1}{3} \right)
     \left( 11 + 6 \ln\frac{\mu_W^2}{M_W^2} \right) \,,\CL
  C_\pm^{(2)}(\mu_W) 
  =& -\frac{1}{3600} (135677 \mp 124095) \CL
   &+ \frac{1}{18} (7 \pm 51) \pi^2
     \mp \frac{1}{2} \left(1 \mp \frac{1}{3} \right) T(x_t) \CL
   & \hspace{-2pc}
     -\frac{5}{36} (11 \mp 249) \ln\frac{\mu_W^2}{M_W^2}
     +\frac{1}{6} (7 \pm 51) \ln^2\frac{\mu_W^2}{M_W^2} \,,
\end{align}
where
\begin{align}
  T(x_t) 
  =& \frac{112}{9} + 32 x_t + \left( \frac{20}{3} + 16 x_t \right) \ln x_t \CL
   &- (8 + 16 x_t) \sqrt{4 x_t -1} 
   \mathrm{Cl}_2 \left( 2 \arcsin\left(\frac{1}{2\sqrt{x_t}}\right) \right) \,, 
\end{align}
$x_t = m_t^2(\mu_W)/M_W^2$, and $\mathrm{Cl}_2(x)$ is given in Eq.~\eqref{eq:cl2}.

In numerical evaluation, we use an approximated number which is
obtained using \texttt{Mathematica},
\begin{align}
  \label{eq:numeric_consts2}
  &\mathrm{Cl}_2 \left( 2 \arcsin\left(\frac{1}{2\sqrt{x_t}}\right) \right)
  = 0.8464504 \ldots
\end{align}
The value $x_t$ is evaluated with the top quark mass
$m_t(m_t)=163.3\;\GeV$ and $M_W=80.385\;\GeV$, approximating
$m_t(\mu_W)=m_t(m_t)$.

Running from $\mu_W$ to the bottom quark threshold $\mu_b$ is achieved
by
\begin{align}
  \label{eq:running1}
  C_{i}(\mu_b,5) 
  =& K_i(\mu_b,5) 
  \left( \frac{\alpha_s(\mu_W,5)}{\alpha_s(\mu_b,5)} \right)^{a_i(5)}
    K_i^{-1}(\mu_W,5) C_{i}(\mu_W) \CL
  =& C_i^{(0)}(\mu_b,5) 
   + \frac{\alpha_s(\mu_b,5)}{4\pi} C_i^{(1)}(\mu_b,5) \CL
   &+ \left(\frac{\alpha_s(\mu_b,5)}{4\pi}\right)^2 C_i^{(2)}(\mu_b,5) \,.
\end{align}

The threshold correction at $\mu_b$ is given by the following.
Writing
\begin{align}
  C_i (\mu_b,4)
  =& C_i^{(0)}(\mu_b,4) 
    + \frac{\alpha_s(\mu_b,4)}{4\pi} C_i^{(1)}(\mu_b,4) \CL
   &+ \left(\frac{\alpha_s(\mu_b,4)}{4\pi}\right)^2 C_i^{(2)}(\mu_b,4) \,,
 \end{align}
then
\begin{align}
  \label{eq:decoupling1}
  C_i^{(0)} (\mu_b,4) =& C_i^{(0)} (\mu_b,5) \,,\CL
  C_i^{(1)} (\mu_b,4) =& C_i^{(1)} (\mu_b,5) \,,\CL
  C_i^{(2)} (\mu_b,4) =& C_i^{(2)} (\mu_b,5) - \delta C_i^{(2)} (\mu_b) \,,
\end{align}
where
\begin{align}
  &\delta C_\pm^{(2)} (\mu_b)
  =- \frac{2}{3} \ln \frac{\mu_b^2}{m_b^2} C_\pm^{(1)}(\mu_b,5) \CL
   &- \left( \frac{2}{3} \ln \frac{\mu_b^2}{m_b^2} r_\pm^{(1)}(\mu_b,5)
     + \delta r_\pm^{(2)}(\mu_b) \right) C_\pm^{(0)}(\mu_b,5) \CL
   &=- \frac{2}{3} \ln \frac{\mu_b^2}{m_b^2} C_\pm^{(1)}(\mu_b,5) \CL
   &+ \left( \pm \left(1 \mp \frac{1}{3} \right) 
       \left( \frac{59}{36} + \frac{1}{3} \ln \frac{\mu_b^2}{m_b^2}
       + \ln^2 \frac{\mu_b^2}{m_b^2} \right)\right) C_\pm^{(0)}(\mu_b,5)  \,.
\end{align}
The definition of $r_i^{(1)}$ and $\delta r_i^{(2)}$, and their
combination, the multiplicative factor of $C_i^{(0)}$, can be found in
Ref.~\cite{ Buras2006:JHEP.11.002}.

Running from $\mu_b$ to the matching scale $\mu_c = m_c(m_c)$ is
achieved by
\begin{align}
  \label{eq:running2}
  C_{i}(\mu_c,4) 
  =& K_i(\mu_c,4) 
  \left( \frac{\alpha_s(\mu_b,4)}{\alpha_s(\mu_c,4)} \right)^{a_i(4)}
    K_i^{-1}(\mu_b,4) C_{i}(\mu_b,4) \CL
  =& C_i^{(0)}(\mu_c,4) 
   + \frac{\alpha_s(\mu_c,4)}{4\pi} C_i^{(1)}(\mu_c,4) \CL
   &+ \left(\frac{\alpha_s(\mu_c,4)}{4\pi}\right)^2 C_i^{(2)}(\mu_c,4) \,.
\end{align}

In the matching calculation, we need the expansion in the three flavor
strong coupling, Eq.~\eqref{eq:expandCif3}.
Equating $C_i(\mu_c)$ in Eq.~\eqref{eq:expandCif3} to $C_i(\mu_c,4)$ after applying Eq.~\eqref{eq:alpha_s_threshold} to the flavor threshold with $f=4$, then
\begin{align}
  \label{eq:Cil_convert}
  C_i^{(0)}(\mu_c) 
  =& C_i^{(0)}(\mu_c,4) \,, \CL
  C_i^{(1)}(\mu_c) 
  =& C_i^{(1)}(\mu_c,4) \,, \CL
  C_i^{(2)}(\mu_c) 
  =& C_i^{(2)}(\mu_c,4) 
    + \frac{2}{3} \ln \frac{\mu_c^2}{m_c^2} C_i^{(1)}(\mu_c,4) \,.
\end{align}
Hence, at the matching scale of the charm quark mass $\mu_c=m_c(m_c)$,
we obtain
\begin{equation}
  \label{eq:running_result}
  C_i^{(l)}(m_c) = C_{i}^{(l)}(m_c,4) \,,\quad (l=0,1,2) \,.
\end{equation}

The QCD beta function expansion coefficients $\beta_i$ to four-loop
order are \cite{ Buras2006:JHEP.11.002,
  Chetyrkin2000:CompPhysComm.133.43}:
\begin{align}
  \label{eq:qcdbeta}
  \beta_0 
  =& 11 - \frac{2}{3}f \,,\CL
  \beta_1 
  =& 102 - \frac{38}{3}f \,,\CL
  \beta_2 
  =& \frac{2857}{2} - \frac{5033}{18}f + \frac{325}{54}f^2 \,,\CL
  \beta_3
  =& \frac{149753}{6} + 3564\zeta_3 
      - \left(\frac{1078361}{162} + \frac{6508}{27}\zeta_3\right)f \CL
   &+ \left(\frac{50065}{162} + \frac{6472}{81}\zeta_3\right)f^2
      + \frac{1093}{729}f^3 \,.
\end{align}

The NNNLO decoupling relation of the strong coupling constant at a flavor threshold $\mu$ is \cite{Chetyrkin2000:CompPhysComm.133.43}
\begin{align}
  \label{eq:alpha_s_threshold}
  &\frac{\alpha_s(\mu,f-1)}{4\pi} \CL
  =& \frac{\alpha_s(\mu,f)}{4\pi}
    - \left( \frac{\alpha_s(\mu,f)}{4\pi} \right)^2 
      \frac{2}{3} \ln \frac{\mu^2}{m_h^2} \CL
   &+ \left( \frac{\alpha_s(\mu,f)}{4\pi} \right)^3
      \left( \frac{22}{9} - \frac{38}{3} \ln \frac{\mu^2}{m_h^2}
        + \frac{4}{9} \ln^2 \frac{\mu^2}{m_h^2} 
      \right) \CL
   &+ \left( \frac{\alpha_s(\mu,f)}{4\pi} \right)^4
      \bigg( \frac{564731}{1944} - \frac{2633}{486} (f-1)
        - \frac{82043}{432} \zeta_3 \CL
        &\quad+ \frac{1}{27} \ln \frac{\mu^2}{m_h^2} (-6793 + 281(f-1)) \CL
        &\quad- \frac{131}{9} \ln^2 \frac{\mu^2}{m_h^2}
        - \frac{8}{27} \ln^3 \frac{\mu^2}{m_h^2} 
      \bigg) \,,
\end{align}
where $m_h = m_h(m_h)$ is the scale invariant $\overline{\text{MS}}$
mass of the heavy flavor which is removed from an effective theory
below the threshold $\mu$, and $\zeta_3$ is a Riemann zeta function,
Eq.~\eqref{eq:zetan}.

The running quark mass $m_q(\mu)$, an $\overline{\text{MS}}$ mass at
scale $\mu$, for a fixed number of active flavors $f$ is \cite{
  Chetyrkin2000:CompPhysComm.133.43}
\begin{equation}
  \label{eq:running_m_q}
  \frac{m_q(\mu)}{m_q(\mu_0)}
  = \frac{R(\alpha_s(\mu)/4\pi)}{R(\alpha_s(\mu_0)/4\pi)} \,,
\end{equation}
with
\begin{align}
  R(x) 
  =& x^{c_0} \Bigg\{ 1 + (c_1 - b_1c_0)x \CL
   &+ \frac{1}{2} \left[ (c_1 - b_1c_0)^2
      + c_2 - b_1c_1 + b_1^2c_0 - b_2c_0 \right]x^2 \CL
   &+ \bigg[ \frac{1}{6} (c_1 - b_1c_0)^3 \CL
   &\quad+ \frac{1}{2} (c_1 - b_1c_0)(c_2 - b_1c_1 + b_1^2c_0 - b_2c_0) \CL
   &\quad+ \frac{1}{3} (c_3 -b_1c_2 + b_1^2c_1 - b_2c_1 - b_1^3c_0 \CL
   &\quad+ 2b_1b_2c_0 - b_3c_0)\bigg]x^3 \Bigg\} \,,
\end{align}
where $m_q(\mu_0)$ is the scale invariant mass $m_q = m_q(m_q)$,
$b_i=\beta_i/\beta_0$, and $c_i=\gamma_m^{(i)}/\beta_0$.
The QCD beta function coefficients $\beta_i$ are given in
Eq.~\eqref{eq:qcdbeta}.
The mass anomalous dimensions $\gamma_m^{(i)}$ are known up to
four-loop order,
\begin{align}
  \gamma_m^{(0)} 
  =& 4 \,,\CL
  \gamma_m^{(1)} 
  =& \frac{202}{3} - \frac{20}{9}f \,,\CL
  \gamma_m^{(2)}
  =& 1249 - \left( \frac{2216}{27} + \frac{160}{3}\zeta_3 \right)f
  - \frac{140}{81}f^2 \,,\CL
  \gamma_m^{(3)}
  =& \frac{4603055}{162} + \frac{135680}{27}\zeta_3 - 8800\zeta_5 \CL
   &+ \left(-\frac{91723}{27} - \frac{34192}{9}\zeta_3 + 880\zeta_4
      + \frac{18400}{9}\zeta_5 \right)f \CL
   & \hspace{-2pc}
      + \left(\frac{5242}{243} + \frac{800}{9}\zeta_3 - \frac{160}{3}\zeta_4
      \right)f^2 
      + \left(-\frac{332}{243} + \frac{64}{27}\zeta_3 \right)f^3
      \,.
\end{align}
In numerical evaluation, we use approximated numbers for the Riemann
zeta functions $\zeta_n$ which are obtained using
\texttt{Mathematica},
\begin{align}
  \label{eq:numeric_consts3}
  \zeta_5 =& 1.036927 \ldots \,,
\end{align}
and $\zeta_3$ and $\zeta_4$ are given in Eq.~\eqref{eq:numeric_consts}.

We used Eq.~\eqref{eq:running_m_q} to expand the charm quark mass
about $m_c=m_c(m_c)$ with $f=4$.

Here, we will give numerical results for an initial scale $\mu_W =
80\;\GeV$ and a varying charm scale $\mu_c$, $1.0 \leq \mu_c \leq
2.0\;\GeV$.
To examine the dependence on the scale $\mu_W$, we repeat the same
analysis with $\mu_W = 40, 160\;\GeV$.
The dependence on $\mu_b$ and $m_t(m_t)$ is ignored
\cite{Brod2011:PhysRevLett.108.121801}.
Fig.~\ref{fig:etacc} summarizes these results.
The following are kept fixed for all analyses: the gauge boson masses
$M_Z = 91.1876\;\GeV \,, M_W = 80.385\;\GeV$; quark masses $m_t(m_t) =
163.3\;\GeV \,, m_b(m_b) = 4.163\;\GeV \,, m_c(m_c) = 1.279\;\GeV$;
bottom quark threshold $\mu_b = 5.0\;\GeV$; and the strong coupling
constant that provides an initial value for the running formula,
$\alpha_s(M_Z,5) = 0.1184\;\GeV$.

At the scales $\mu_W = 80\;\GeV$ and $\mu_c = 1.279\;\GeV$,
\begin{align}
  \label{eq:compare}
  \eta_{cc}/[\alpha_s(\mu_c,3)]^{a_+} 
  =& 1.129757 + 0.571608 + 0.430890 \,,\CL
  \eta_{cc}^\text{NNLO} =& 1.738396 \,,\CL
  \eta_{cc}^\text{NLO} =& 1.387098 \,.
\end{align}
The value of $\eta_{cc}^\text{NLO}$ is obtained by summing the first
two terms in the series, and the value of $\eta_{cc}^\text{NNLO}$ is
obtained by summing all three terms in the series.

At the scales $\mu_W = 80\;\GeV$ and $\mu_c = 1.300\;\GeV$,
\begin{align}
  \label{eq:compare}
  \eta_{cc}/[\alpha_s(\mu_c,3)]^{a_+} 
  =& 1.113769 + 0.568911 + 0.433783 \,,\CL
  \eta_{cc}^\text{NNLO} =& 1.720690 \,,\CL
  \eta_{cc}^\text{NLO} =& 1.368023 \,.
\end{align}

We claim the NNLO $\eta_{cc}$ is
\begin{equation}
  \label{eq:yc_etacc}
  \eta_{cc}^\text{NNLO} = 1.72(27) \,.
\end{equation}
The central value corresponds to the result with the scales
$\mu_c=1.3\;\GeV$ and $\mu_W=80\;\GeV$.
For the error, we add the $\mu_c$ and $\mu_W$ dependences in
quadrature,
\begin{align}
  \delta_{\mu_c}
  =& \eta_{cc}^\text{NNLO}(\mu_c=1.3\;\GeV,\mu_W=80\;\GeV) \CL
   &\quad- \eta_{cc}^\text{NNLO}(\mu_c=1.8\;\GeV,\mu_W=80\;\GeV) \CL
  =& 0.266 \,,\CL
  \delta_{\mu_W}
  =& (\eta_{cc}^\text{NNLO}(\mu_c=1.3\;\GeV,\mu_W=160\;\GeV) \CL
   &\quad- \eta_{cc}^\text{NNLO}(\mu_c=1.3\;\GeV,\mu_W=40\;\GeV))/2 \CL
  =& 0.047 \,.
\end{align}
Though we include errors from the inputs $\alpha_s(M_Z,5)$ and
$m_c(m_c)$, the final errors in Eq.~\eqref{eq:yc_etacc} are not
affected; we use the errors quoted in Ref.~\cite{
  Brod2011:PhysRevLett.108.121801}:
\begin{equation}
  \delta_{\alpha_s} = 0.06 \,,\quad
  \delta_{m_c} = 0.01 \,.
\end{equation}

In Ref.~\cite{Brod2011:PhysRevLett.108.121801}, the authors also added
the absolute shift from the NLO value of $\eta_{cc}$.
It is the main reason for their larger error,
\begin{equation}
  \eta_{cc}^\text{NNLO} = 1.87(76) \,.
\end{equation}
The main concern for adding this shift is poor convergence of the
$\alpha_s$ expansion for $\eta_{cc}$.
We, however, take the view that the error from $\mu_c$ dependence
suffices to estimate the size of higher order corrections.

Here, we would like to comment on our choice of $1.3 \, \GeV \le \mu_c
\le 1.8 \, \GeV$.
If we examine each order in the perturbative corrections of
$\eta_{cc}$,
\begin{align}
  \eta_{cc} &= 1_\text{(LO)} + 0.37_\text{(NLO)} + 0.36_\text{(NNLO)}
  \nonumber \\
  & = 1.72 \qquad \text{(SWME)} \,,
  \\
  \eta_{cc} &= 1_\text{(LO)} + 0.38_\text{(NLO)} + 0.49_\text{(NNLO)}
  \nonumber \\
  & = 1.87 \qquad \text{(Brod \& Gorbahn)} \,.  
\end{align}
The error that we quote is 0.27, which is roughly equivalent to the
size of the NNLO correction.\footnote{ If we use the size of the NLO
  correction given in Ref.~\cite{ Buras2008:PhysRevD.78.033005}, the
  NNLO correction is 0.29, which is roughly equivalent to our error
  estimate $\approx 0.27$. }
If we take the interval $1\,\GeV \le \mu_c \le 2 \,\GeV$ as suggested
in Ref.~\cite{Brod2011:PhysRevLett.108.121801}, the error becomes
0.42, which is larger than the NNLO correction.
If we follow the procedure of Ref.~\cite{
  Brod2011:PhysRevLett.108.121801}, we add the NNLO size of 0.36 to
this in quadrature which leads to a total error of 0.55.
This value is significantly larger than the size of the NNLO
correction.
This estimate becomes even larger than that of the NLO and NNLO
corrections combined in quadrature.
If we assume that perturbation theory is working, then it is already
arguably conservative to choose the size of the NNLO correction as the
systematic error due to truncated NNNLO terms.
Hence, we believe that the error quoted in Ref.~\cite{
  Brod2011:PhysRevLett.108.121801} is somewhat overestimated.

In addition, in Ref.~\cite{ Buras2013:EurPhysJC.73.2560}, Buras and
Girrbach suggested that, if one chooses the interval $1.3 \, \GeV \le
\mu_c \le 1.8 \, \GeV$, one can obtain their result for $\eta_{cc}$,
\begin{equation}
  \eta_{cc} \approx 1.70(21) \,,
\end{equation}
which is obtained indirectly through an estimate of the long distance
contribution to $\Delta M_K$ based on a large $N$ QCD inspired model
calculation.
We have directly verified their claim.
%


Our result for $\eta_{cc}^\text{NLO}$ agrees with the value quoted in
Ref.~\cite{ Brod2011:PhysRevLett.108.121801}
\begin{equation}
  \eta_{cc}^\text{NLO} = 1.38(52)(07)(02) \,.
\end{equation}

\section{$\epsK^\text{SM}$ with the SWME $\BK$}
\label{app:epsKwSWME}

Lattice results for the exclusive $V_{cb}$ \cite{
  Bailey2014:PhysRevD.89.114504} and the SWME $\BK$ \cite{
  Bae2014:prd.89.074504} are obtained using overlapping subsets of the
MILC asqtad gauge ensembles \cite{Bazavov:2009bb}.
This implies that there exists a complicated, non-trivial correlation
between them.
It is possible to calculate, in principle, this correlation exactly
from the data set.
Unfortunately, this correlation is not available yet.
Hence, the current situation is that we need to estimate the
systematic error due to the unknown correlation between $\BK$ and
$V_{cb}$.

Here is our strategy.
First, we take the uncorrelated case as the central value.
Second, we introduce $+50\%$ correlation between $\BK$ and
$V_{cb}$ and obtain results for $\epsK^\text{SM}$.
Third, we introduce $-50\%$ anti-correlation between $\BK$ and
$V_{cb}$ and repeat the analysis to obtain $\epsK^\text{SM}$.
Fourth, we take the maximum deviation from the central value as the
systematic error due to the unknown correlation between $\BK$ and
$V_{cb}$.

\begin{figure*}[!t]
  \subfigure[CKMfitter]{%
    \includegraphics[width=0.31\textwidth]{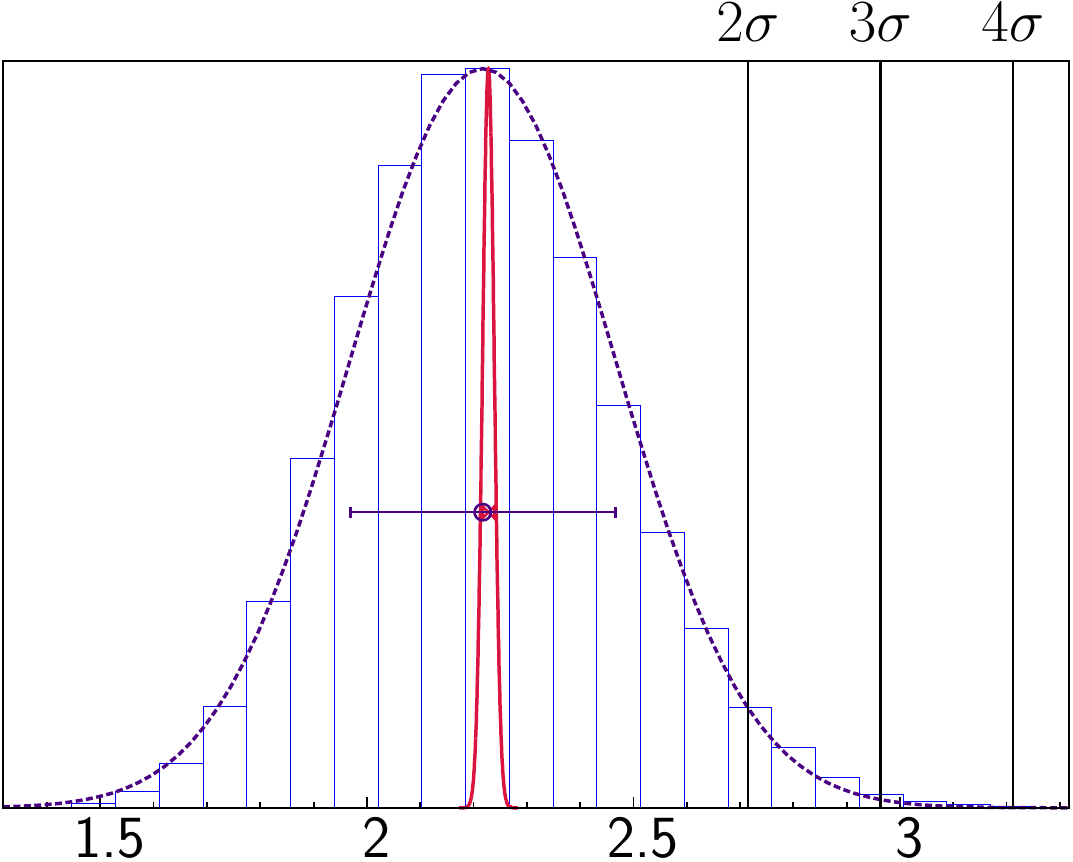}
    \label{sfig:incl.SWME.CKM}
  }
  \hfill
  \subfigure[UTfit]{%
    \includegraphics[width=0.31\textwidth]{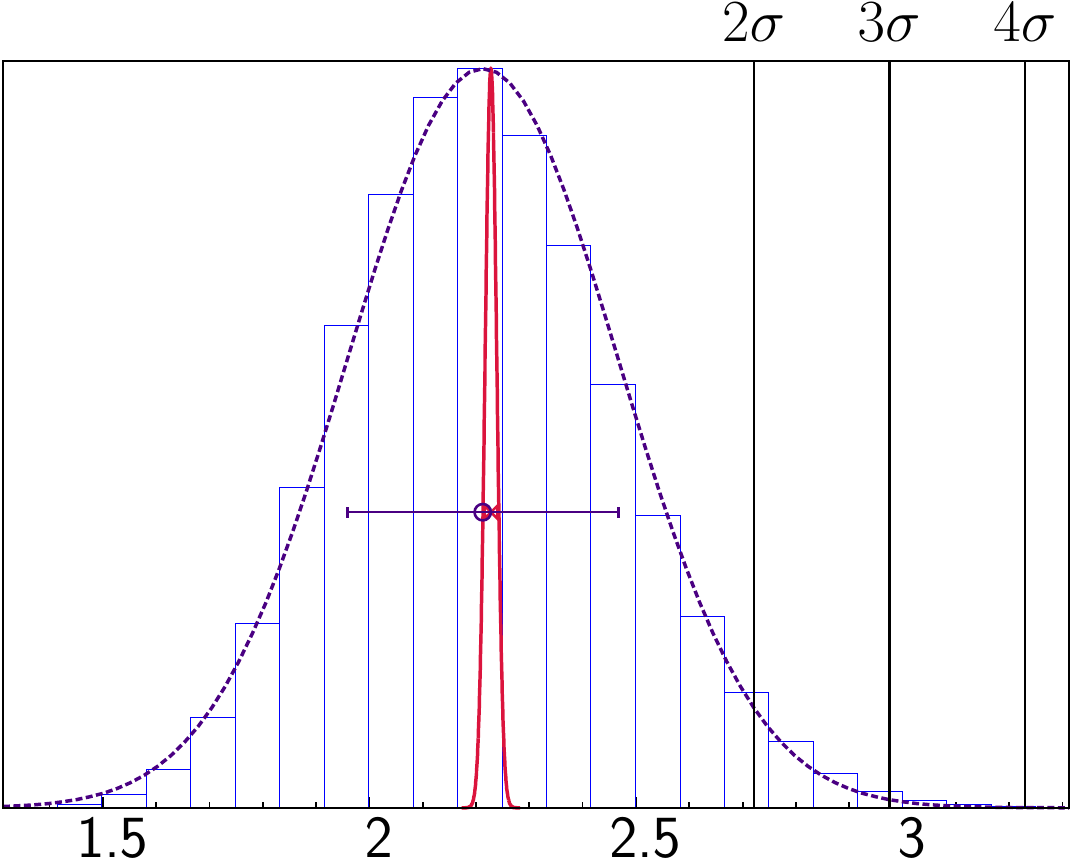}
    \label{sfig:incl.SWME.UT}
  }
  \hfill
  \subfigure[AOF]{%
    \includegraphics[width=0.31\textwidth]{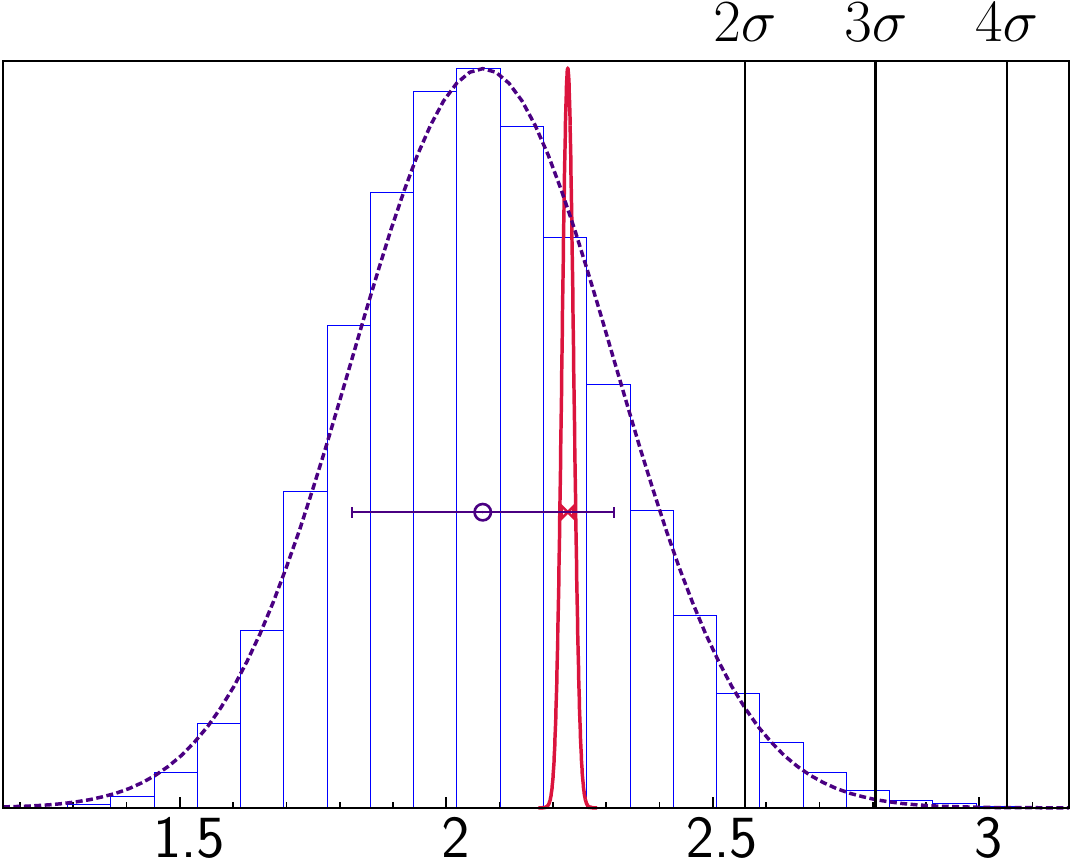}
    \label{sfig:incl.SWME.AO}
  }
  \\
  \subfigure[CKMfitter]{%
    \includegraphics[width=0.31\textwidth]{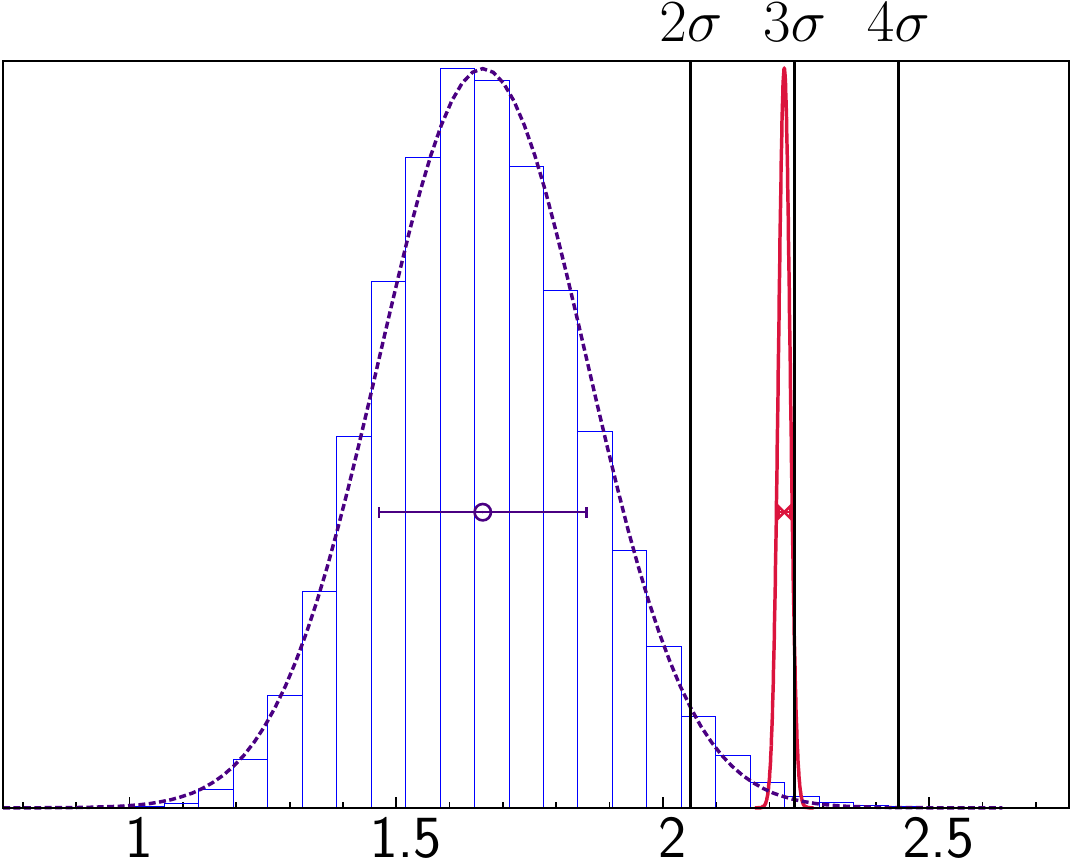}
    \label{sfig:excl.SWME.CKM}
  }
  \hfill
  \subfigure[UTfit]{%
    \includegraphics[width=0.31\textwidth]{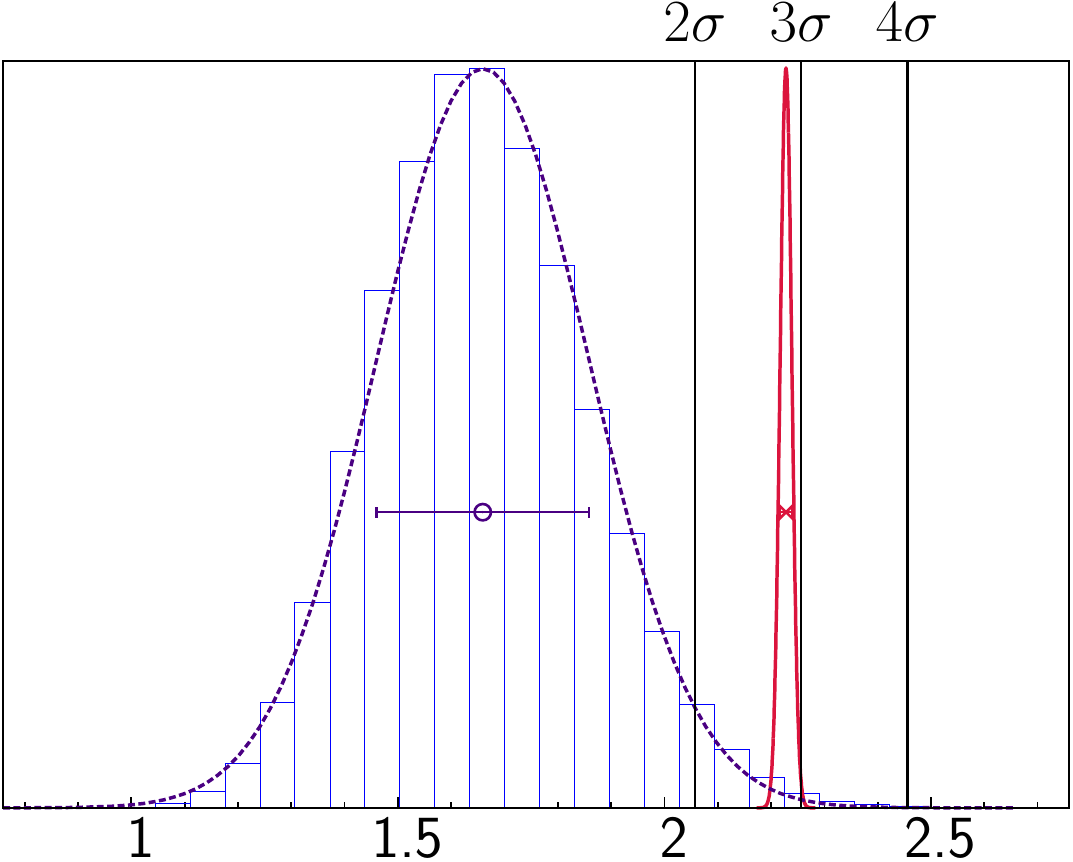}
    \label{sfig:excl.SWME.UT}
  }
  \hfill
  \subfigure[AOF]{%
    \includegraphics[width=0.31\textwidth]{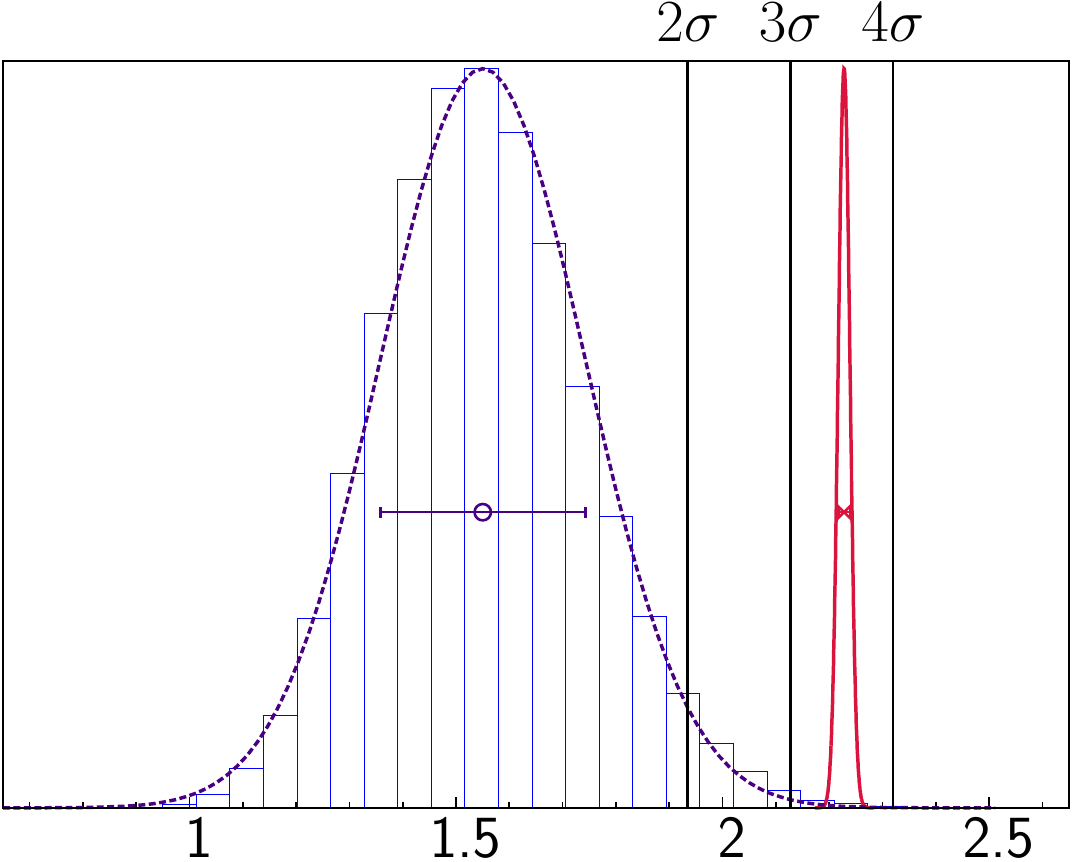}
    \label{sfig:excl.SWME.AO}
  }
  \caption{Gaussian probability distributions for $\epsK^\text{SM}$
    (blue dotted line) and $\epsK^\text{Exp}$ (red solid line) with
    the SWME $\BK$.  Here, we assume no correlation between $\BK$ and
    $V_{cb}$.  Results of \ref{sfig:incl.SWME.CKM},
    \ref{sfig:incl.SWME.UT} and \ref{sfig:incl.SWME.AO} are obtained
    using the inclusive $\Vcb$.  Results of \ref{sfig:excl.SWME.CKM},
    \ref{sfig:excl.SWME.UT} and \ref{sfig:excl.SWME.AO} are obtained
    using the exclusive $\Vcb$.  }
  \label{fig:hstgwSWME}
\end{figure*}

In Table \ref{tbl:epsKwSWME}, we present results for $\epsK^\text{SM}$
for the uncorrelated case.
In Table \ref{tbl:DepsKwSWME}, we present the corresponding results
for $\Delta \epsK$.
In Fig.~\ref{fig:hstgwSWME}, we show the corresponding probability
distribution of $\epsK^\text{SM}$.
\begin{table}[!h]
  \caption{ $\epsK^\text{SM}$ in units of $10^{-3}$.  We use the SWME
    $\BK$ with no correlation between $\BK$ and $V_{cb}$. }
  \label{tbl:epsKwSWME}
  \renewcommand{\arraystretch}{1.2}
  \begin{ruledtabular}
  \begin{tabular}{ccc}
  Input Method & Inclusive $\Vcb$ & Exclusive $\Vcb$ 
  \\ \hline
  CKMfitter 
  & $2.22(25)$ 
  & $1.66(19)$ 
  \\ \hline
  UTfit
  & $2.21(25)$ 
  & $1.66(20)$ 
  \\ \hline
  AOF
  & $2.07(25)$ 
  & $1.55(19)$ 
  \end{tabular}
  \end{ruledtabular}
\end{table}
\begin{table}[!h]
  \caption{ $\Delta \epsK$ with no correlation between $\BK$ and
    $V_{cb}$. We take $\epsK^\text{SM}$ from
    Table~\ref{tbl:epsKwSWME}. $\epsK^\text{Exp}$ is given in
    Eq.~\eqref{eq:epsK_exp}.}
  \label{tbl:DepsKwSWME}
  \renewcommand{\arraystretch}{1.2}
  \begin{ruledtabular}
  \begin{tabular}{ccc}
  Input Method & Inclusive $\Vcb$ & Exclusive $\Vcb$ 
  \\ \hline
  CKMfitter 
  & $0.04\sigma$ 
  & $2.9\sigma$ 
  \\ \hline
  UTfit
  & $0.06\sigma$ 
  & $2.9\sigma$ 
  \\ \hline
  AOF
  & $0.65\sigma$ 
  & $3.5\sigma$ 
  \end{tabular}
  \end{ruledtabular}
\end{table}

\begin{figure*}[!t]
  \subfigure[CKMfitter]{%
    \includegraphics[width=0.31\textwidth]{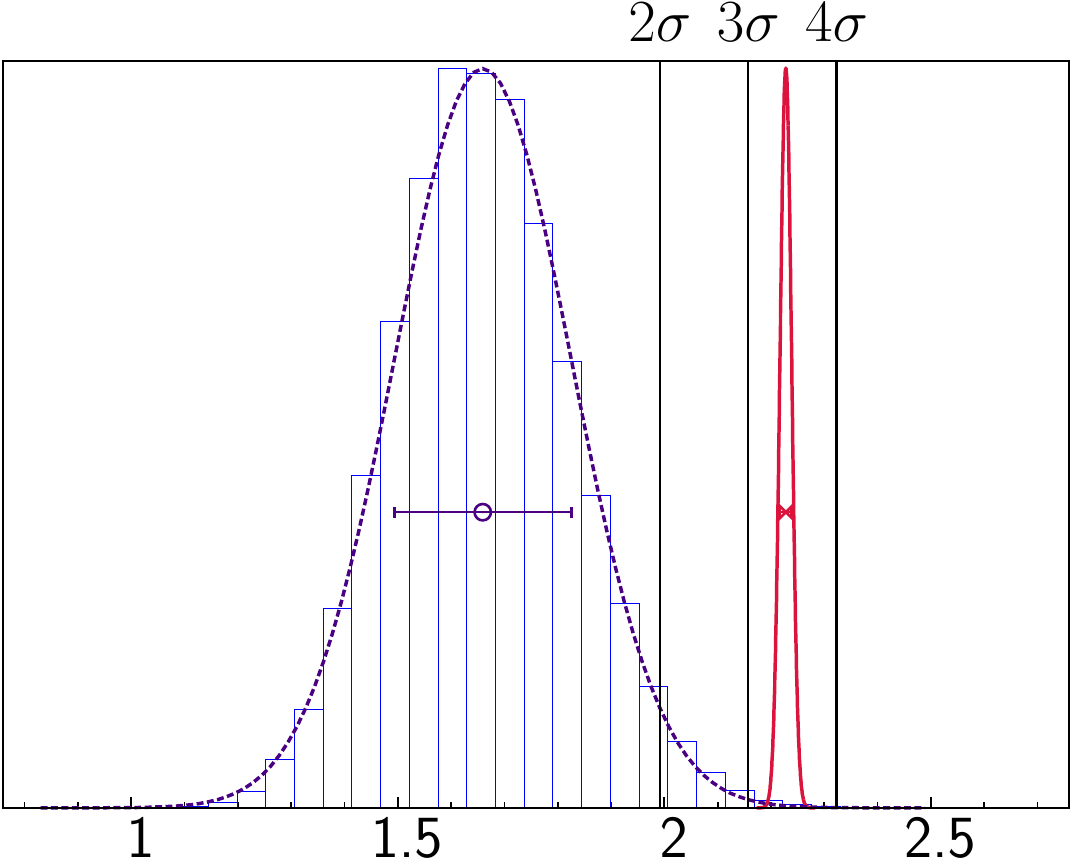}
    \label{sfig:excl.SWME.CKM.nc50}
  }
  \hfill
  \subfigure[UTfit]{%
    \includegraphics[width=0.31\textwidth]{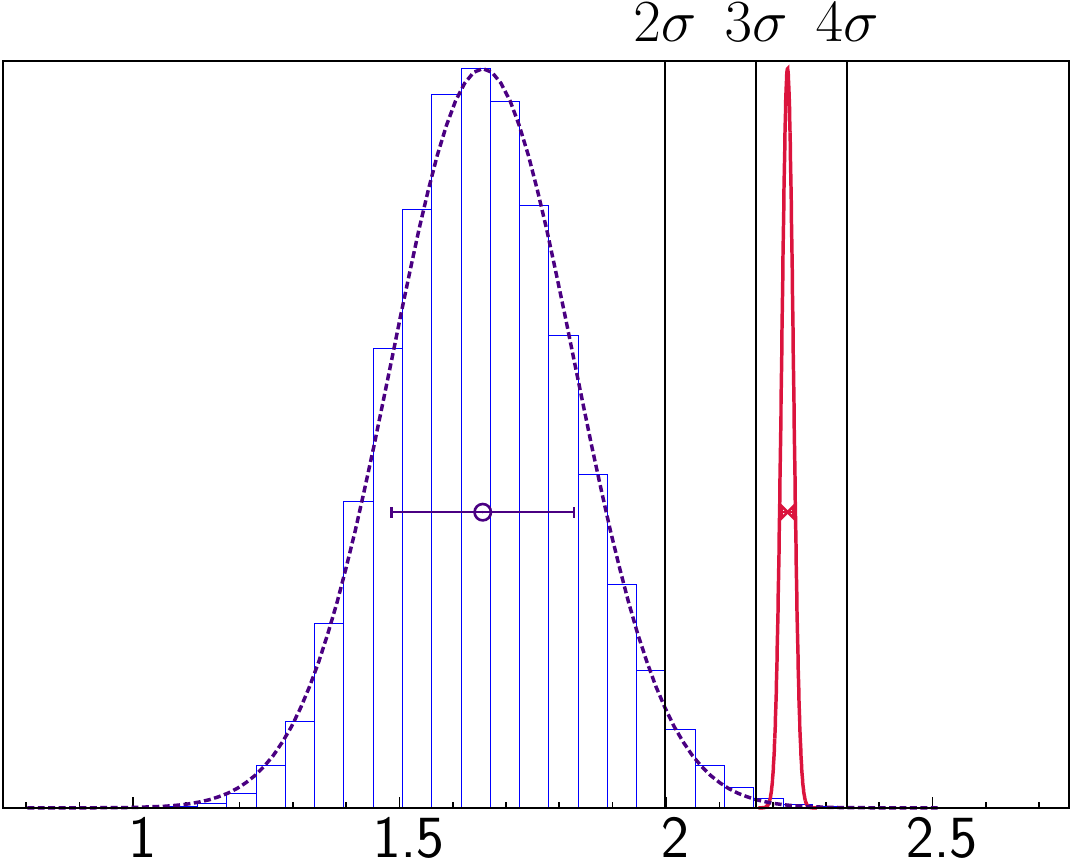}
    \label{sfig:excl.SWME.UT.nc50}
  }
  \hfill
  \subfigure[AOF]{%
    \includegraphics[width=0.31\textwidth]{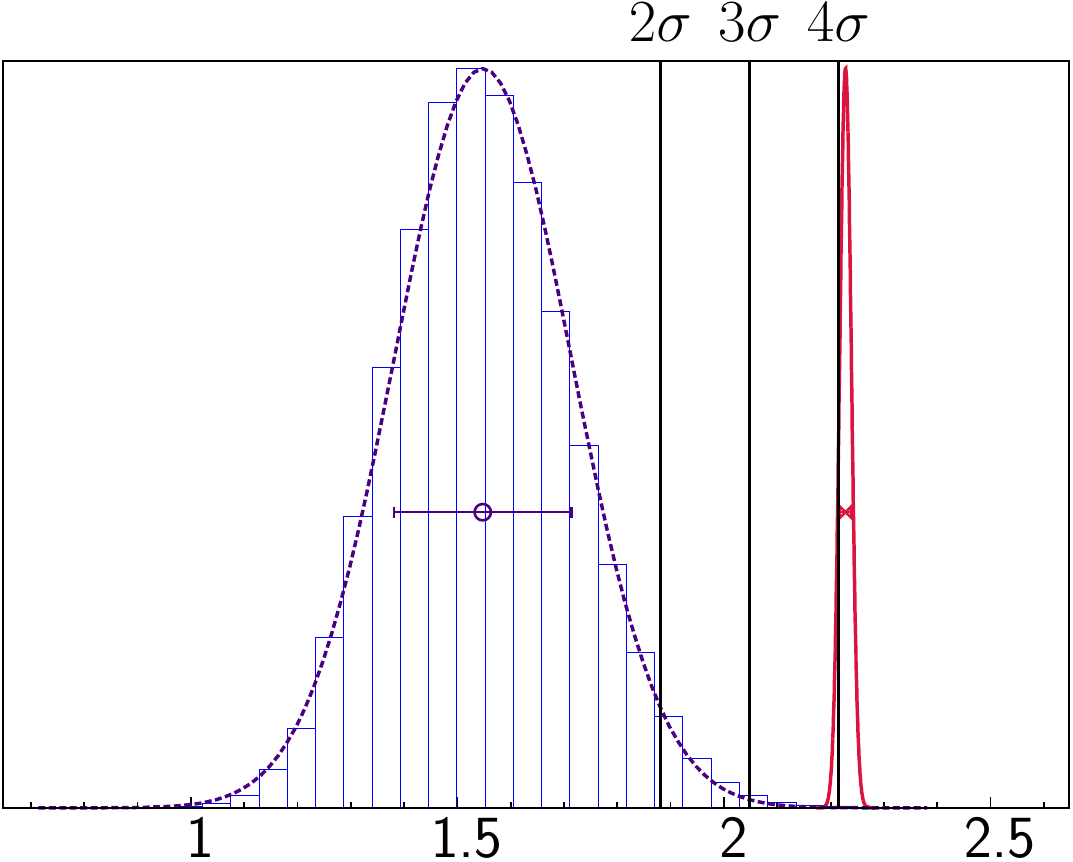}
    \label{sfig:excl.SWME.AO.nc50}
  }
  \\
  \subfigure[CKMfitter]{%
    \includegraphics[width=0.31\textwidth]{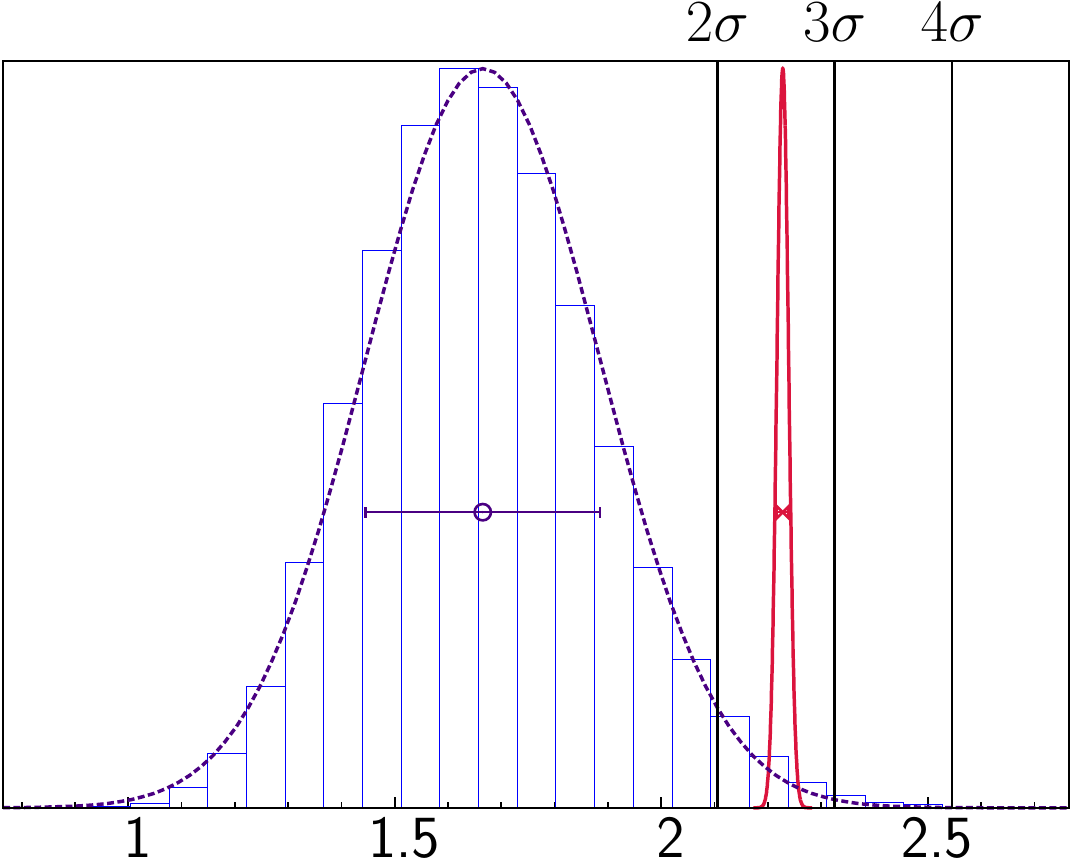}
    \label{sfig:excl.SWME.CKM.pc50}
  }
  \hfill
  \subfigure[UTfit]{%
    \includegraphics[width=0.31\textwidth]{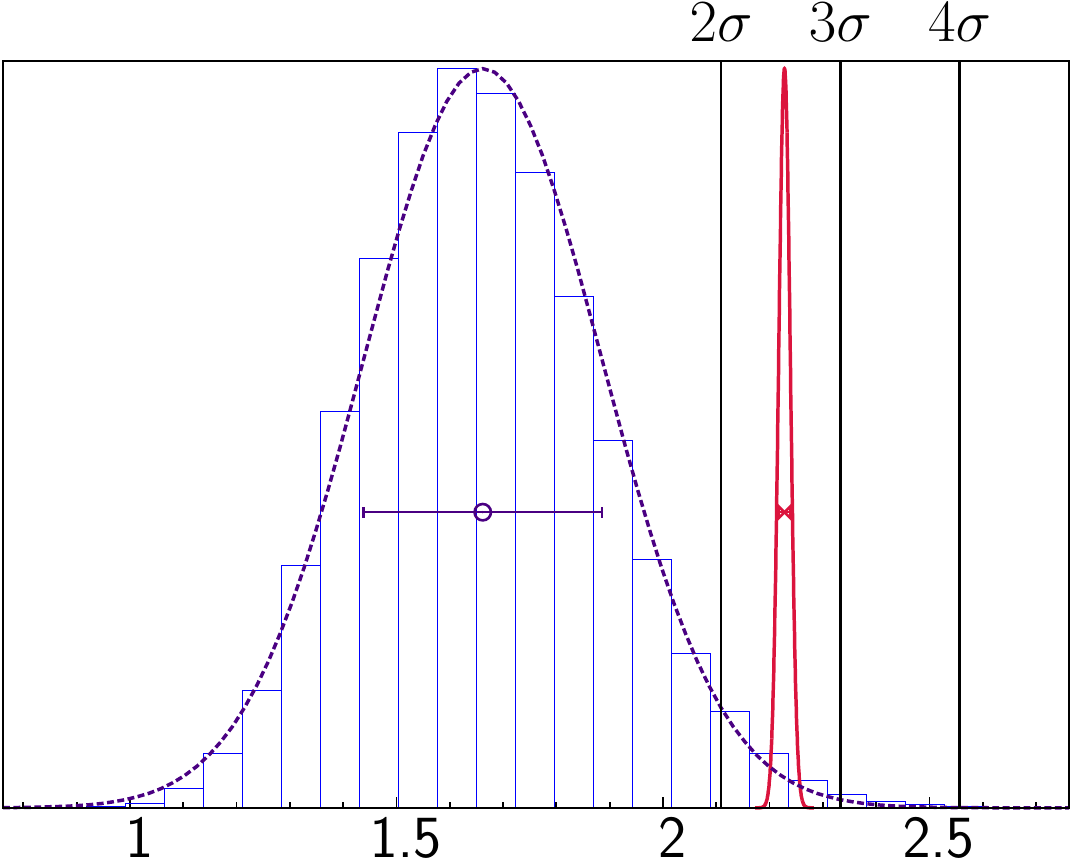}
    \label{sfig:excl.SWME.UT.pc50}
  }
  \hfill
  \subfigure[AOF]{%
    \includegraphics[width=0.31\textwidth]{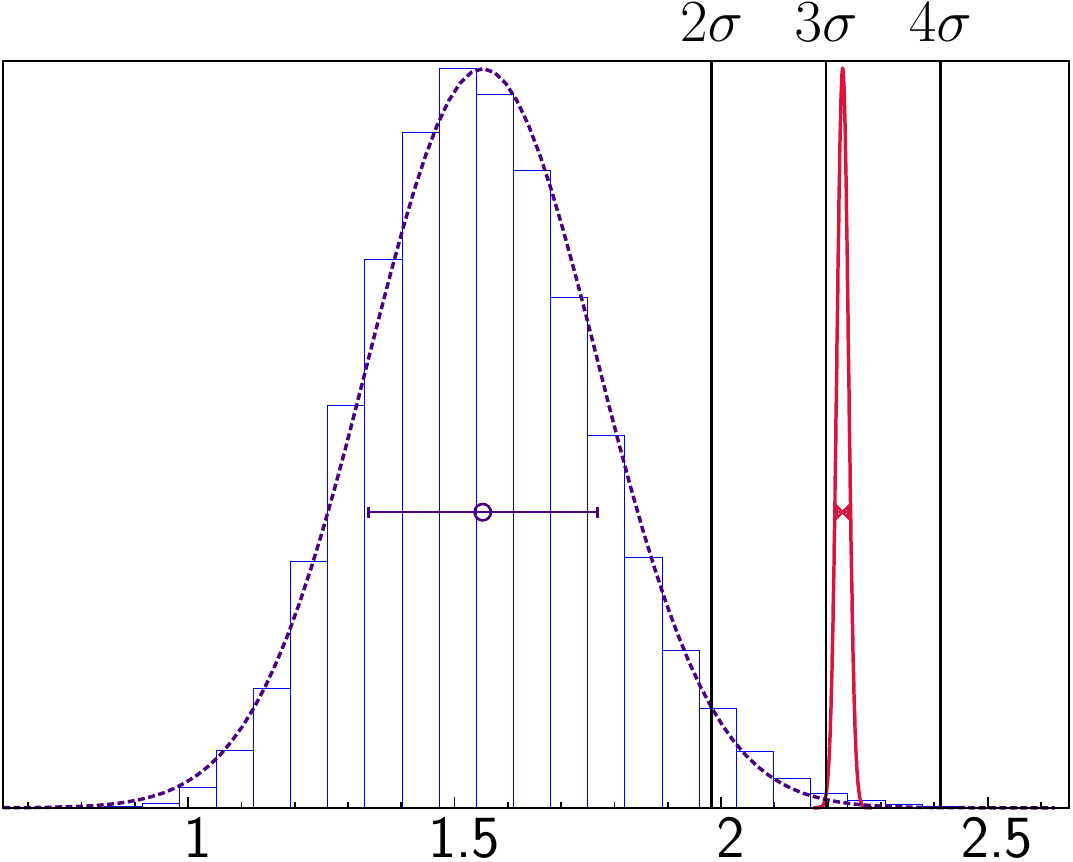}
    \label{sfig:excl.SWME.AO.pc50}
  }
  \caption{Gaussian probability distributions for $\epsK^\text{SM}$
    and $\epsK^\text{Exp}$ with the SWME $\BK$ and exclusive $\Vcb$.
    Results of \ref{sfig:excl.SWME.CKM.nc50},
    \ref{sfig:excl.SWME.UT.nc50}, and \ref{sfig:excl.SWME.AO.nc50} are
    obtained with $-50\%$ anti-correlation between $\BK$ and
    $V_{cb}$. Those of \ref{sfig:excl.SWME.CKM.pc50},
    \ref{sfig:excl.SWME.UT.pc50}, and \ref{sfig:excl.SWME.AO.pc50} are
    obtained with $+50\%$ correlation between $\BK$ and $V_{cb}$.  }
  \label{fig:hstgwSWMEwCorr}
\end{figure*}

In Table \ref{tbl:epsKwSWMEwCorr}, we present results for
$\epsK^\text{SM}$ with $+50\%$ correlation and $-50\%$
anti-correlation between $\BK$ and (exclusive) $V_{cb}$.
In Table \ref{tbl:DepsKwSWMEwCorr}, we present the corresponding
results for $\Delta \epsK$.
In Fig.~\ref{fig:hstgwSWMEwCorr}, we show the probability distribution
for the corresponding $\epsK^\text{SM}$.
\begin{table}[!h]
  \caption{$\epsK^\text{SM}$ in units of $10^{-3}$.  We use the SWME
    $\BK$ and the exclusive $\Vcb$ with $+50\%$ correlation and
    $-50\%$ anti-correlation between them.}
  \label{tbl:epsKwSWMEwCorr}
  \renewcommand{\arraystretch}{1.2}
  \begin{ruledtabular}
  \begin{tabular}{ccc}
  Input Method & $c=-50\%$ & $c=+50\%$ 
  \\ \hline
  CKMfitter 
  & $1.66(17)$ 
  & $1.67(22)$ 
  \\ \hline
  UTfit
  & $1.66(17)$ 
  & $1.66(22)$ 
  \\ \hline
  AOF
  & $1.55(17)$ 
  & $1.55(21)$ 
  \end{tabular}
  \end{ruledtabular}
\end{table}
\begin{table}[!h]
  \caption{$\Delta \epsK$ with $+50\%$ correlation and $-50\%$
    anti-correlation between $\BK$ and exclusive $V_{cb}$. We take
    $\epsK^\text{SM}$ from
    Table~\ref{tbl:epsKwSWMEwCorr}. $\epsK^\text{Exp}$ is given in
    Eq.~\eqref{eq:epsK_exp}.}
  \label{tbl:DepsKwSWMEwCorr}
  \renewcommand{\arraystretch}{1.2}
  \begin{ruledtabular}
  \begin{tabular}{ccc}
  Input Method & $c=-50\%$ & $c=+50\%$ 
  \\ \hline
  CKMfitter 
  & $3.4\sigma$ 
  & $2.6\sigma$ 
  \\ \hline
  UTfit
  & $3.3\sigma$ 
  & $2.5\sigma$ 
  \\ \hline
  AOF
  & $4.1\sigma$ 
  & $3.1\sigma$ 
  \end{tabular}
  \end{ruledtabular}
\end{table}

Hence, we obtain the final results for $\Delta \epsK$ for the SWME
$\BK$ and exclusive $V_{cb}$:
\begin{align}
  \Delta \epsK &= (3.5 \pm 0.6) \sigma\,, 
  \label{eq:DepsK-SWME}
\end{align}
where the error represents the uncertainty due to the
correlation between $\BK$ and exclusive $V_{cb}$.

First, the results in Eq.~\eqref{eq:DepsK-SWME} are consistent with
those in Eq.~\eqref{eq:final-DepsK:ex} within the systematic errors.
Second, the correlation between $\BK$ and exclusive $V_{cb}$ 
dominates the error in $\Delta \epsK$ with the SWME $\BK$.
In addition, this error is much larger than that in our final results
in Eq.~\eqref{eq:final-DepsK:ex}.
Hence, we use the results with the SWME $\BK$ only to cross-check
those with the FLAG $\BK$.


\bibliographystyle{apsrev} 
\bibliography{ref} 
\end{document}

%% file: macro.tex
\providecommand{\wbar}[1]{\overline#1}
\providecommand{\abs}[1]{\lvert#1\rvert}

\providecommand{\ket}[1]{\lvert#1\rangle}
\providecommand{\mate}[3]{\langle#1\lvert#2\rvert#3\rangle}
\providecommand{\expv}[1]{\langle#1\rangle}

\renewcommand{\Re}{\mathrm{Re}}
\renewcommand{\Im}{\mathrm{Im}}

\providecommand{\GeV}{\mathrm{GeV}}

\providecommand{\CL}{\nonumber\\}

\definecolor{HLBlue}{HTML}{6599FF}
\definecolor{HLOrange}{HTML}{FF6600}

\newcommand{\BK}{\hat{B}_{K}}
\newcommand{\Vcb}{V_{cb}}

\newcommand{\eps}{\varepsilon}
\newcommand{\teps}{\tilde{\varepsilon}}
\newcommand{\epsK}{\varepsilon_{K}}

\newcommand{\org}[1]{\textcolor{gray}{#1}} 